\documentclass[aps,prc, nofootinbib,
preprintnumbers,showkeys, reprint,
superscriptaddress]{revtex4-1}

\usepackage{graphicx}
\usepackage{tikz}
\usepackage{multirow}
\usetikzlibrary{shapes.misc}

\tikzset{
    cross/.pic = {
    \draw[rotate = 45, very thick] (-#1,0) -- (#1,0);
    \draw[rotate = 45, very thick] (0,-#1) -- (0, #1);
    }
}
\usepackage{physics}
\usetikzlibrary{decorations.markings}
\usepackage{subcaption}
\usepackage{dcolumn}
\usepackage{bm}
\usepackage{amsmath}
\usepackage{wasysym}
\usepackage{float}
\usepackage{multirow}
\usepackage{xcolor}
\usepackage{scrextend}
\usepackage[colorlinks=true, 
pdfstartview=FitV, 
bookmarks=true, 
bookmarksnumbered=true, 
breaklinks]{hyperref}
\usepackage{color}
\usepackage[normalem]{ulem} 
\definecolor{blue}{rgb}{0.0, 0.0, 1.0}
\definecolor{red}{rgb}{1.0, 0.0, 0.0}
\definecolor{royalblue}{rgb}{0.0, 0.14, 0.4}
\hypersetup{linkcolor=royalblue, 
	citecolor=blue, 
	urlcolor=royalblue}

\def\orcid#1{\kern .08em\href{https://orcid.org/#1}{\includegraphics[keepaspectratio,width=0.7em]{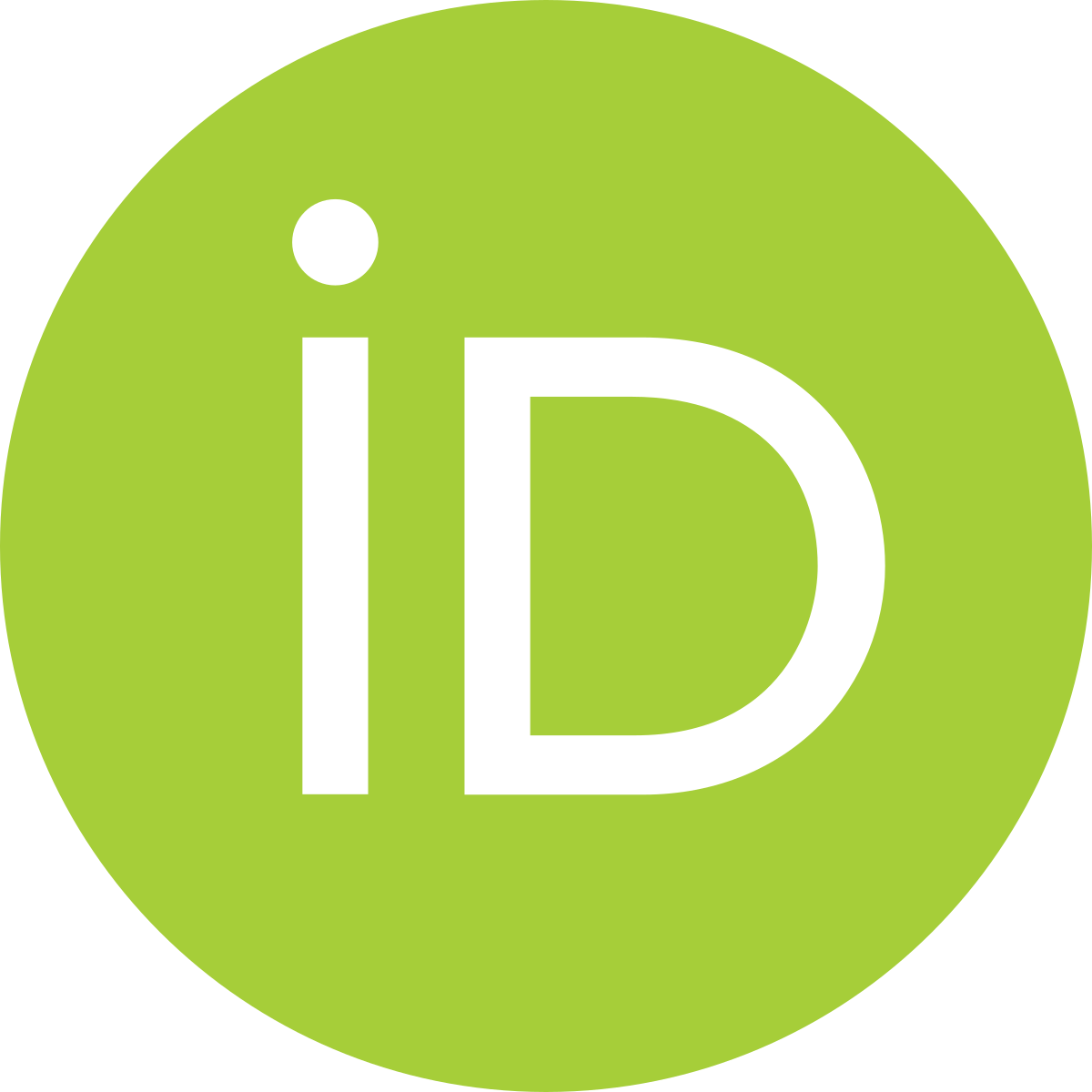}}}

\definecolor{ttcolor}{RGB}{240,248,255}
\definecolor{btcolor}{RGB}{144,238,144}
\definecolor{tbcolor}{RGB}{216,191,216}
\definecolor{bbcolor}{RGB}{255,228,225}

\usepackage{soul}
\usepackage{booktabs}

\begin{document}
\title{A Deep Learning Framework for Disentangling Triangle Singularity 
	\\ and Pole-Based Enhancements}

\author{Darwin Alexander O. Co\orcid{0009-0003-8982-8727}}
\email[]{doco@up.edu.ph}
\affiliation{National Institute of Physics, University of the Philippines Diliman, Quezon City 1101, Philippines}

\author{Vince Angelo A. Chavez\orcid{0009-0009-2373-1985}}
\email[]{vachavez@up.edu.ph}
\affiliation{National Institute of Physics, University of the Philippines Diliman, Quezon City 1101, Philippines}

\author{Denny Lane B. Sombillo\orcid{0000-0001-9357-7236}}
\email[Corresponding author: ]{dbsombillo@up.edu.ph}
\affiliation{National Institute of Physics, University of the Philippines Diliman, Quezon City 1101, Philippines}

\date{\today}
\begin{abstract}
    \noindent
    Enhancements in the invariant mass distribution or scattering cross-section are usually associated with resonances. However, the nature of exotic signals found near hadron-hadron thresholds remain a puzzle today due to the presence of experimental uncertainties. In fact, a purely kinematical triangle diagram is also capable of producing similar structures, but do not correspond to any unstable quantum state. In this paper, we report for the first time, that a deep neural network can be trained to distinguish triangle singularity from pole-based enhancements with a reasonably high accuracy of discrimination between the two seemingly identical line shapes. We also identify the type of triangle enhancement that can be misidentified as a dynamic pole structure. We apply our method to confirm that the $P_\psi^N(4312)^+$ state is not due to a triangle singularity, but is more consistent with a pole-based interpretation, as determined solely through pure line-shape analysis. Lastly, we explain how our method can be used as a model-selection framework useful in studying other exotic hadron candidates.
\end{abstract}



\maketitle

\section{Introduction}\label{sec:intro}
The internal structure of hadrons and their interactions are governed by quantum chromodynamics (QCD) in the low-energy regime. In this energy region, the perturbative technique is unreliable due to large coupling strength. One possible direction to make progress is to construct a hadronic spectrum from experimental observations. Through the spectrum, one can identify the excited states of ground-state hadrons which can give us a hint of how quarks and gluons arrange themselves. This approach of utilizing a bottom-top analysis is fully described in \cite{JPAC:2021rxu}. Typically, excited states of hadrons manifest as peaks in the scattering cross-section or invariant mass distribution. However, not all peaks or structures in the experimental data fit the description of a compact hadronic resonance \cite{Olsen:2017bmm,Guo:2017jvc}. For example, peaks appearing just below a two-hadron threshold are naturally interpreted as hadronic molecules \cite{Hanhart:2014ssa, Hyodo:2014bda}. These peaks are produced mainly due to the coupled-channel effects, appearing as resonance-like structures in the lower mass channel. Hadronic molecular signals, though not part of the true hadronic spectrum of excited hadrons, are useful to pinpoint how hadrons interact among themselves. It is therefore useful to distinguish between compact resonance signals and hadronic molecular signals.

In addition to compact and hadronic molecules, some signals do not correspond to unstable quantum states. These signals are generally referred to as cusps and are typically associated with the kinematics of reactions \cite{Guo:2019twa}. Due to the branch-point singularity associated with the opening of a new channel, one expects a sudden change in the line shape of an $s$-wave amplitude at the threshold. This change in the line shape is further enhanced by the presence of poles located even in the distant Riemann sheet. The problem with cusps is that they can be misinterpreted as quantum states due to the uncertainty present in experimental data. It therefore warrants more careful analysis to discriminate them from true resonances \cite{Guo:2014iya}. Distinguishing different origins of signals, especially near thresholds, is one of the long-standing issues in hadron spectroscopy.

One of the novel approaches in analyzing line shapes is through the use of machine learning techniques \cite{JPAC:2021rxu,He:2023zin,Boehnlein:2021eym}. In particular, the problem of line shape interpretation can be phrased as a classification problem where the input corresponds to line shape and the output is the interpretation or the origin of enhancements. Once the classification problem is unambiguously defined, a deep neural network (DNN) can be trained to map the input line shape space to the output interpretation space. This mapping is made possible because a trained DNN can be considered as a universal approximator \cite{Hornik:1989yye,Schmidhuber:2015ysx}. Such line shape analysis framework was first utilized in \cite{Sombillo:2020ccg} where a neural network was developed to distinguish between a bound and a virtual state enhancement at the nucleon-nucleon threshold. The method was later applied to the pion-nucleon system in \cite{Sombillo:2021rxv} where a DNN is trained to identify the distribution of poles in different Riemann sheets. The expected next step is to use the machine learning technique to analyze exotic signals. This attempt was first done in \cite{Ng:2021ibr} where a DNN was trained using a coupled-channel effective range approximation to identify the origin of the $P_{\psi}^{N}(4312)^{+}$  signal. It was further demonstrated in the same study that the machine learning technique conforms with the result of the conventional analysis made in \cite{Fernandez-Ramirez:2019koa}. Further investigations of other exotic signals using machine learning are done in \cite{Liu:2022uex,Zhang:2023czx,Chen:2022ddj}.

Existing machine learning studies have focused entirely on the conventional interpretations of exotic signals and have not yet considered the possible role of kinematical effects. One cannot simply rule out the kinematic interpretation when it has been shown to also produce resonance-like enhancements near two-hadron thresholds \cite{Guo:2019twa,Guo:2014iya}. A more general approach would be to train a DNN to recognize and distinguish line shapes arising from both the dynamic pole structure and the kinematical effects in a reaction \cite{Liu:2015taa}. This way, it can be used to identify the origin of signals without any particular biases. One such kinematical mechanism prominently used in the study of exotic signals is the triangle singularity.

\begin{figure}[ht!]
	\centering
	\includegraphics[width=0.8\linewidth]{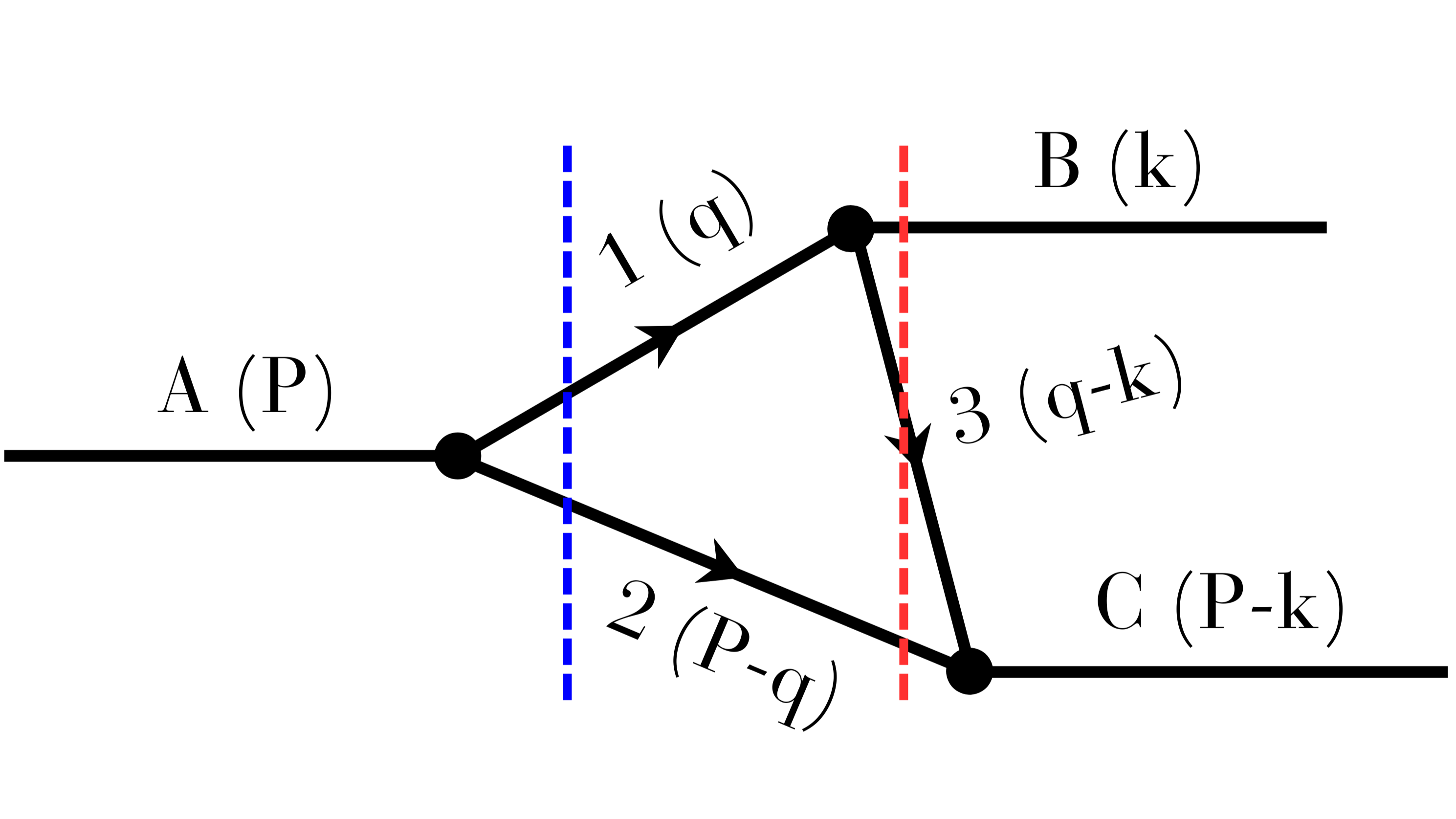} 
	\caption[]{Triangle diagram for a general reaction $A\to B + C$, where 1, 2, and 3 are the on-shell intermediate particles. The quantities enclosed in parentheses are the corresponding momenta of the particles. Figure reconstructed from \cite{Guo:2019twa}.}
	\label{fig:triangle_diagram}
\end{figure}

First proposed by Landau \cite{Landau:1959fi}, the triangle singularity is a kinematic re-scattering process involving three intermediate particles, such that the decay amplitude has a singularity close to the physical region which produces an enhancement in the invariant mass distribution of the final state \cite{Guo:2019twa}. This process is represented by a triangle diagram, illustrated in Fig.~\ref{fig:triangle_diagram}, involving an initial state $A$ decaying into final states $B$ and $C$, with intermediate particles 1, 2, and 3. During the process, particle $A$ decays into particles 1 and 2, then particle 1 decays into particles 3 and $B$. Particle 3 then catches up to particle 2 along the same direction and collides to form particle $C$ in the final state. For it to produce enhancements in the physical region, it is required that the interaction at all vertices be classical processes and all the intermediate particles be real propagators---on their mass shell with real momenta \cite{Coleman:1965xm}.

The triangle singularity regained prominence in the community when it was used to explain some newly observed exotic signals \cite{Wu:2011yx,Wu:2012pg, Mikhasenko:2015oxp,Aceti:2016yeb,Szczepaniak:2015eza, Nakamura:2019nwd}. In particular, it was suggested that the exotic hidden-charm pentaquark state $P_c(4450)^+$ observed by the LHCb collaboration in 2015 \cite{LHCb:2015yax} could be the manifestation of a $\Lambda(1890)\chi_{c1}p$ triangle loop in the $\Lambda_b^0 \to J/\psi\,p K^-$ decay \cite{Guo:2015umn,Liu:2015fea,Bayar:2016ftu,Mikhasenko:2015vca}. The renewed interest in studying this mechanism is due to the similarities between the enhancements produced by a triangle singularity and that of a true hadronic resonance, which once again emphasizes the issues in interpreting ambiguous experimental line shapes. Since then, the nature of triangle singularity and its application in analyzing exotic signals has been extensively studied. For a complete review, see \cite{Guo:2019twa}.

In 2019, an updated analysis of the decay was reported by LHCb wherein the narrow pentaquark state $P_\psi^N(4312)^+$ (then called the $P_c(4312)^+$), together with the two-peak structure of the $P_c(4450)$ resonance, was observed with notable statistical significance \cite{LHCb:2019kea}. In their analysis of these signals, the triangle mechanism was ruled out as a reasonable interpretation for the $P_\psi^N(4312)^+$ and $P_c(4440)^+$ states due to lack of proper hadron-rescattering thresholds in their triangle diagrams. It, however, remains a plausible hypothesis for the $P_c(4457)^+$ state, requiring further amplitude analysis to establish a more definitive interpretation. Other studies have also pointed out the possibility of a general double triangle mechanism capable of producing similar structures \cite{Nakamura:2021qvy}.

In this study, we apply machine learning techniques to analyze the various mechanisms relevant to the $P_\psi^N(4312)^+$ state, including the kinematic triangle diagram. We develop, for the first time, a deep neural network capable of distinguishing triangle singularity from pole-based enhancements via pure line-shape analysis. We benchmark our framework by applying the trained DNN on the $P_\psi^N(4312)^+$ state and validate if the triangle singularity interpretation can be consistently ruled out. In doing so, we demonstrate the effectiveness of our novel model-selection framework and its significance as a possible tool for classifying exotic hadron candidates.

The paper is organized as follows: in sections \ref{sec:trianglesingularity} and \ref{sec:uniformization}, we review the formalism of the triangle singularity and the $S$-matrix, respectively, and how they apply to $P_\psi^N(4312)^+$. Section \ref{sec:DNN} discusses the step-by-step construction, design, and training of the deep neural network model. In section \ref{sec:results}, we validate our trained DNN and apply it to interpret $P_\psi^N(4312)^+$. Finally, in section \ref{sec:conclusion}, we report our conclusions and future outlooks.

\section{Triangle Singularity}\label{sec:trianglesingularity}
The analytic structure of an amplitude with triangle singularity (TS) is readily captured in the scalar one-loop integral given by
\begin{widetext}
	\begin{equation}
		I(k)=i\int\dfrac{d^4q}{(2\pi)^4}
		\dfrac{1}{\left(q^2-m_1^2+i\epsilon\right)
			\left[(P-q)^2-m_2^2+i\epsilon\right]
			\left[(q-k)^2-m_3^2+i\epsilon\right]},
		\label{eq:full_I}
	\end{equation}
\end{widetext}
where $P$, $k$, $q$ are the momenta of the particles as labelled in Fig.~\ref{fig:triangle_diagram}. The masses $m_1$, $m_2$, and $m_3$ correspond to the masses of the intermediate particles, and $\epsilon$ pertains to the decay width of the internal diagram.

Since the intermediate particles must be real propagators, one can apply on-shell treatments as discussed in \cite{Guo:2019twa}, and rewrite the integral as
\begin{equation}
	I(k)\propto \int_0^\infty \dfrac{q^2 f(q)\,dq}{P^0-\sqrt{m_1^2+q^2}-\sqrt{m_2^2+q^2}+i\epsilon},
	\label{eq:approx_I}
\end{equation}
where $q=|\vec{q}|$ is now the magnitude of the 3-momentum vector and $P^0$ is the total energy of the initial state. This simplified form is sufficient in the investigation of the underlying singularities that will emerge from the triangle loop. Since particles $m_1$ and $m_2$ are on-shell, the integrand has a singularity at values of $q$ satisfying $P^0=\sqrt{m_1^2+q^2}+\sqrt{m_2^2+q^2}-i\epsilon$, which corresponds to the left blue cut appearing in Fig.~\ref{fig:triangle_diagram}. The function $f(q)$ in equation \eqref{eq:approx_I} is the corresponding angular part of the volume integral given by
\begin{equation}
	\resizebox{0.4\textwidth}{!}{$f(q)=\displaystyle\int_{-1}^{1}\dfrac{dz}{E_C-\omega(q)-\sqrt{m_3^2+q^2+k^2-2qkz}+i\epsilon},$}
	\label{eq:f(q)}
\end{equation}
where $E_C = (m_A^2-m_B^2+m_C^2)/2m_A$ is the energy of particle $C$ in the rest frame of $A$, $\omega(q)=\sqrt{m_2^2+q^2}$, and $k=\sqrt{\lambda(m_A,m_B,m_{23})}/2m_A$, with $\lambda(x,y,z)=x^2+y^2+z^2-2xy-2yz-2xz$. The $m_{23}$ parameter is the invariant mass of the final state. By straightforward substitution, one can calculate the integral in equation \eqref{eq:f(q)} and find that it contains a logarithmic expression given by
\begin{equation}
	\resizebox{0.4\textwidth}{!}{$f(q)=\frac{1}{qk}\left[(u_+ - u_-) - (E_C - \omega(q) + i\epsilon)\ln(\frac{u_+}{u_-})\right],$}
	\label{eq:f(q)_solved}
\end{equation}
where $u_{\pm}=E_C - \omega(q) - \sqrt{m_3^2+q^2+k^2\mp2qk}$. The logarithmic singularity at $q$ satisfies $u_{\pm}=0$, corresponding to the right red cut in Fig.~\ref{fig:triangle_diagram}. 

To ensure the convergence of the integral, we may include a form factor with an expression given by $\Lambda^2/(\Lambda^2+q^2)$, where we define $\Lambda$ as some arbitrary cut-off parameter related to the range of hadronic interaction \cite{Nakamura:2019nwd}. 

Essentially, we find that with equations \eqref{eq:approx_I} and \eqref{eq:f(q)_solved}, the integral $I$ can now be expressed as a simple function of all the involved particle's masses. In other words, the underlying singularities are completely dependent on kinematical variables alone and do not involve any dynamical interactions within the system. It is, however, sensitive to these variables, meaning that the enhancement would not occur if certain kinematic conditions are not satisfied. Still, we have arrived at a useful result that allows us to directly compute for intensity $|I|^2$ and construct the line shape for any triangle diagram once we determine the involved masses.

Further analysis of the kinematical region where the triangle singularity arises leads to an exact relation between the masses in the internal triangle diagram. It is derived and established in \cite{Bayar:2016ftu} that the triangle singularity can only occur within the energy region of
\begin{equation}
	m_{23}^2 \in \left[(m_2+m_3)^2,\;\frac{M_am_3^2-M_b^2m_2}{M_a-m_2}+M_am_2\right],
	\label{eq:m23_range}
\end{equation}
where $M_a$ and $M_b$ are the masses of the external particles $A$ and $B$, respectively. Moreover, we find the relation
\begin{equation}
	m_1^2 \in \left[\frac{M_a^2m_3+M_b^2m_2}{m_2+m_3}-m_2m_3,\;(M_a-m_2)^2\right],
	\label{eq:m1_range}
\end{equation}
which consequently establishes a constraint for the masses allowed in the triangle loop.
These prescribed ranges shall become relevant when choosing the appropriate triangle loops capable of producing resonance-like peaks in a given reaction. Another implication is that one may easily control the position of the resulting peak in the invariant mass by simply adjusting the masses of the intermediate particles. We utilize this as our guiding principle for generating multiple TS line shapes as training data for the DNN. This is discussed in section \ref{sec:DNN}.

The $P_\psi^N(4312)^+$ structure, with a measured signal at $4311.9\pm0.7$ MeV, can be observed just below the $\Sigma_c^+\bar{D}^{*0}$ threshold. From the analysis performed by the LHCb in \cite{LHCb:2019kea}, a triangle loop shown in Fig.~\ref{fig:triangle_diagram}, with internal lines $123$ corresponding to $D_s^{**-}\Lambda_c^+\bar{D}^{*0}$, can be constructed where the mass of some hypothetical $D_s^{**-}$ meson is set to 3288 MeV. This configuration of intermediate particles in the internal lines, with their corresponding masses, gives rise to a line shape that fit the experimental data with a peak at 4312 MeV.

\begin{figure}[h!!!]
	\centering
	\includegraphics[width=\linewidth]{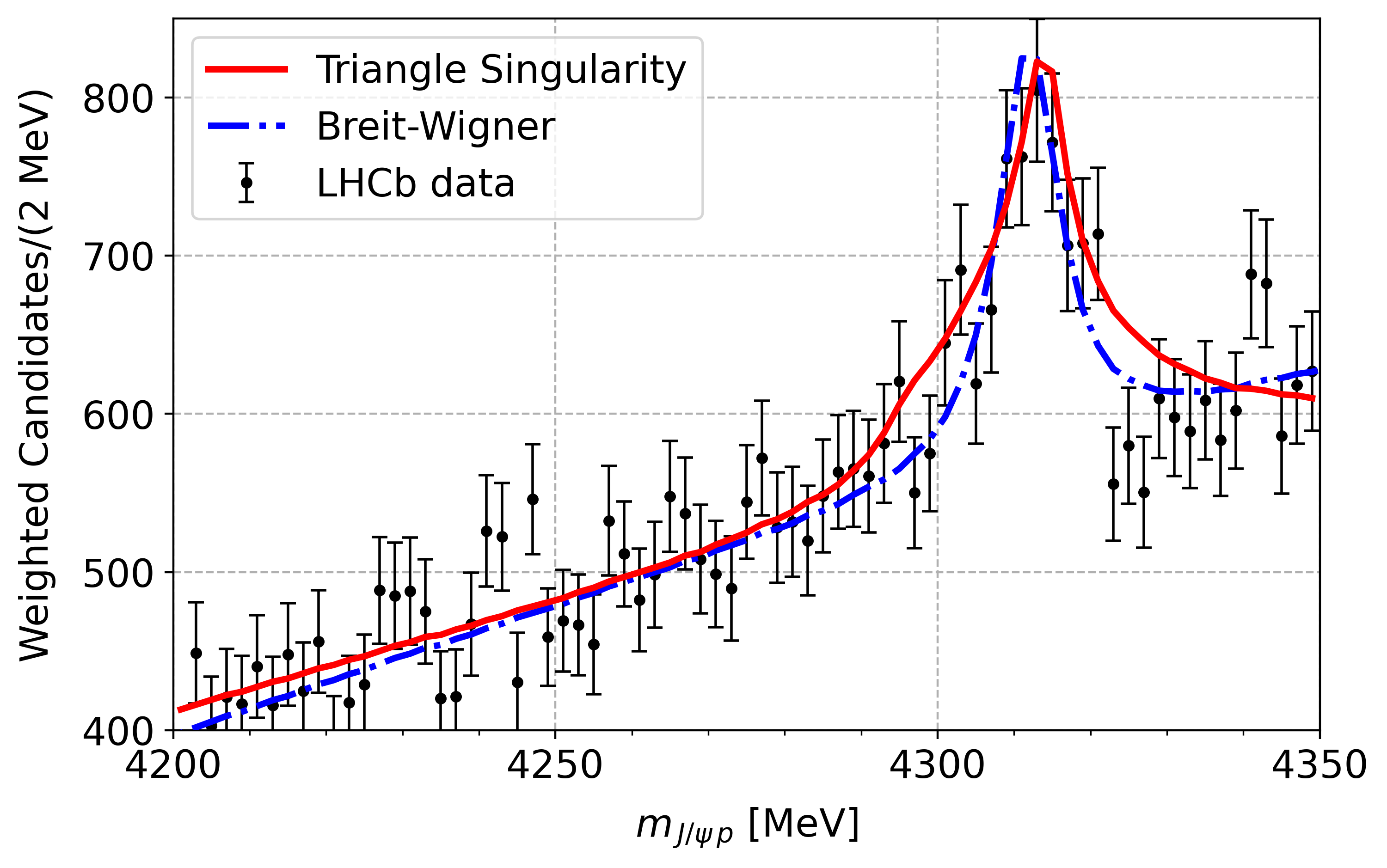}
	\caption[]{The $\cos\theta_{P_c}$-weighted $J/\psi p$ invariant mass distribution with fitted triangle singularity (red) and Breit-Wigner amplitude (blue). LHCb experimental data are retrieved from \cite{hepdata.89271}.}\label{fig:Pc4312_TSBW}
\end{figure}

In order to illustrate the ambiguity arising between a triangle mechanism and a typical resonance enhancement, we plot them simultaneously in Fig.~\ref{fig:Pc4312_TSBW}. Using the Particle Data Group (PDG) values \cite{Workman:2022ynf} of the hadron masses involved in the triangle diagram, one can compute the integral in equation \eqref{eq:approx_I} and plot the $|I|^2$ contribution as a function of the invariant mass of the final state, which in this case is $J/\psi p$. A peak around the 4312 MeV can be produced by setting the values of $\epsilon=1$ MeV and $\Lambda=3000$ MeV to regulate the integral. Care must be taken in considering these parameters since this would suggest that all the particles in the internal lines are stable point hadrons which is not the case. The same argument was used by LHCb to rule out the TS interpretation for the $P^{N}_{\psi}(4312)^+$ state \cite{LHCb:2019kea}. The resonance interpretation can be implemented  by using the Breit-Wigner parametrization to fit the experimental cross-section. The Breit-Wigner amplitude is given by
\begin{equation}
	F(s)\propto\frac{1}{(E^2-M^2)+iM\Gamma},
	\label{eq:BW}
\end{equation} 
where $E$ is the scattering energy, $M$ and $\Gamma$ are the mass and decay width of the unstable state. For both the Breit-Wigner and TS amplitudes, we multiply a projected phase space of the $J/\psi p$ invariant mass and some arbitrary high-order degree polynomial, following the expression
\begin{equation}
	\dfrac{dN}{d\sqrt{s}}=\rho(\sqrt{s})\left[|F(\sqrt{s})|^2+B(\sqrt{s})\right],
	\label{eq:full_amplitude}
\end{equation}
where $\rho(\sqrt{s})$ is the phase space factor, $F(\sqrt{s})$ is the constructed amplitude (either Breit-Wigner or the triangle integral $I$), and $B(\sqrt{s})$ is the additional polynomial background \cite{Fernandez-Ramirez:2019koa}. The $dN/d\sqrt{s}$, corresponding to the resulting intensity, achieves a better fit of the experimental data by optimizing the parameters in $F(\sqrt{s})$ and $B(\sqrt{s})$. The resulting distribution is shown in Fig.~\ref{fig:Pc4312_TSBW} where a narrow peak at around 4312 MeV is produced by both line shapes. Since both mechanisms fit within the uncertainties (error bars) of the experimental data, further analysis is required to unambiguously discriminate between the two largely different mechanisms and interpretations, despite their similar line shapes.

Although this triangle mechanism fits the experimental data well, it was argued to be an unlikely origin of the enhancement due to some physical constraints \cite{LHCb:2019kea}. First, the TS description lacks an appropriate hadron-rescattering threshold, as it requires the exchange of the hypothetical $D_s^{**-}(3288)$ hadron in the loop.  This excited meson has not yet been observed and is not listed in the PDG's compilation of $D_s$ states. Second, the LHCb analysis indicated that achieving a good fit with the TS mechanism necessitates the use of unphysical parameters. Specifically, when the decay width $\epsilon$ is set to some higher and more plausible value, say 50 MeV, the enhancement from the TS is no longer prominent and falls into the background. This physical argument is the main reason why a dynamic pole-based interpretation is favored for the $P_\psi^N(4312)^+$ state.

\begin{figure*}[htb!!]
	\centering
	\begin{subfigure}[b]{0.47\columnwidth}
		\centering
		\includegraphics[scale = 0.1]{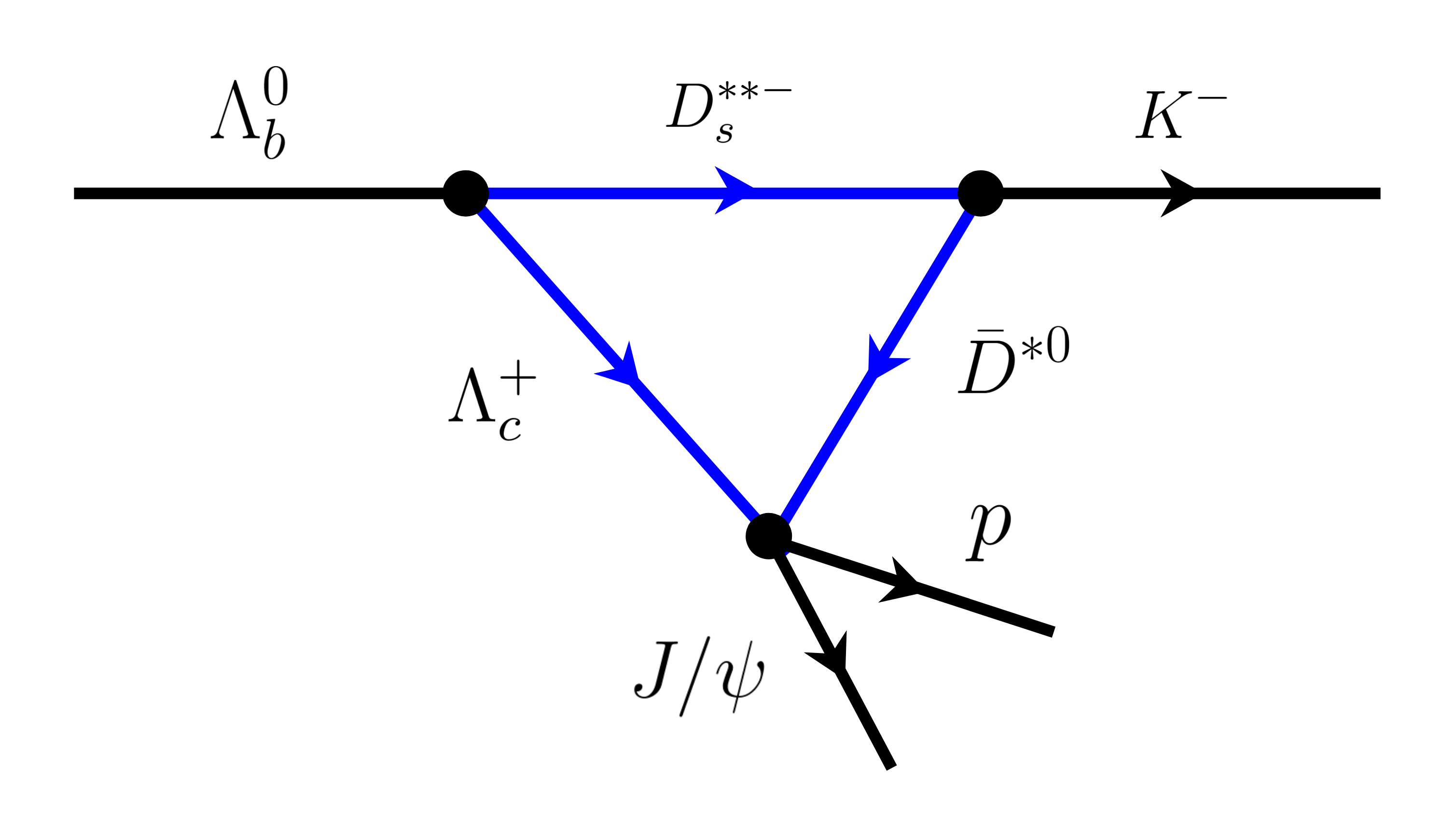}
		\caption[]{ }\label{fig:TSloopA}
		
	\end{subfigure}
	\begin{subfigure}[b]{0.47\columnwidth}  
		\centering 
		\includegraphics[scale = 0.1]{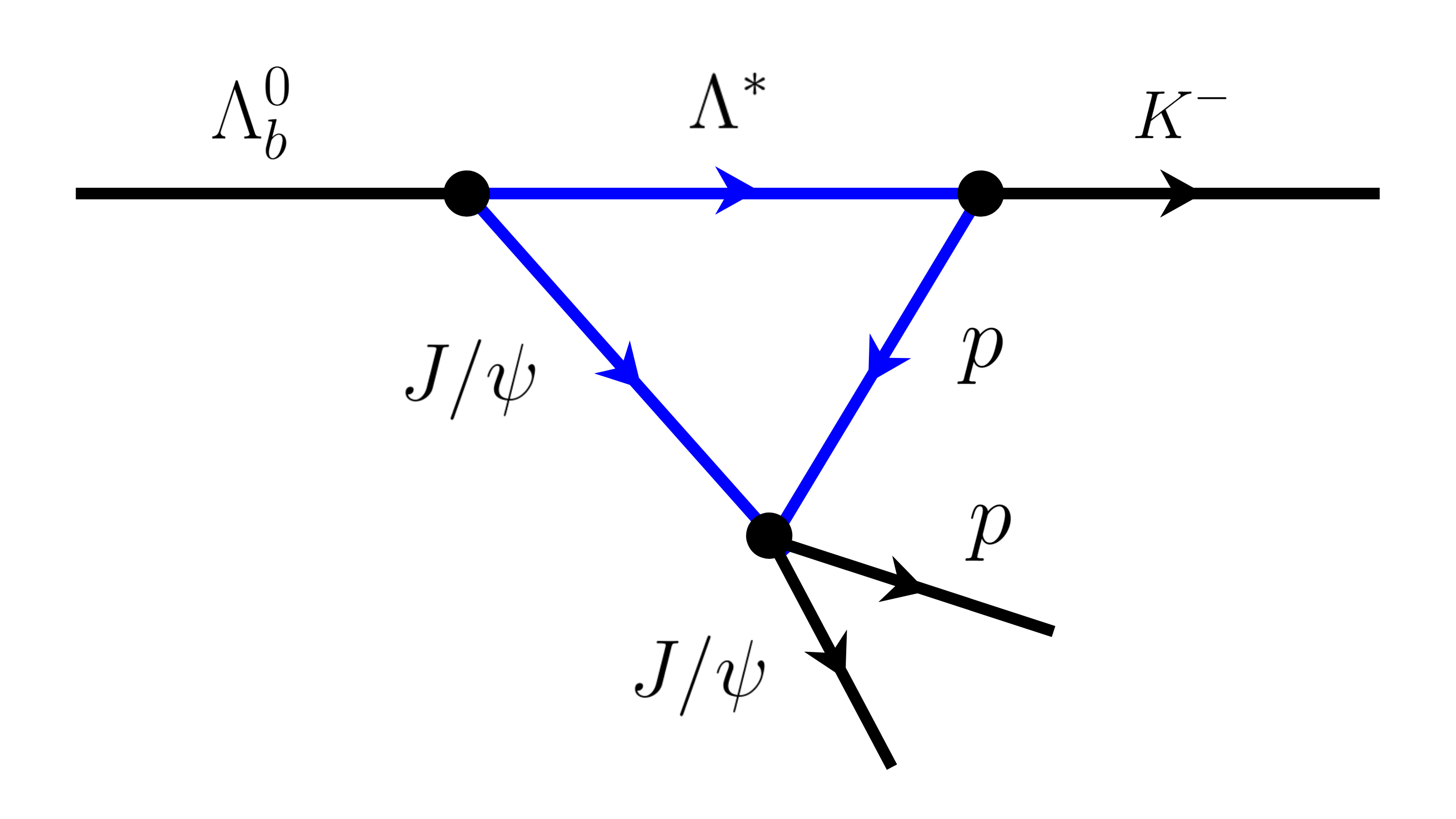}
		\caption[]{ }\label{fig:TSloopB}
		
	\end{subfigure}
	\begin{subfigure}[b]{0.47\columnwidth}   
		\centering 
		\includegraphics[scale = 0.115]{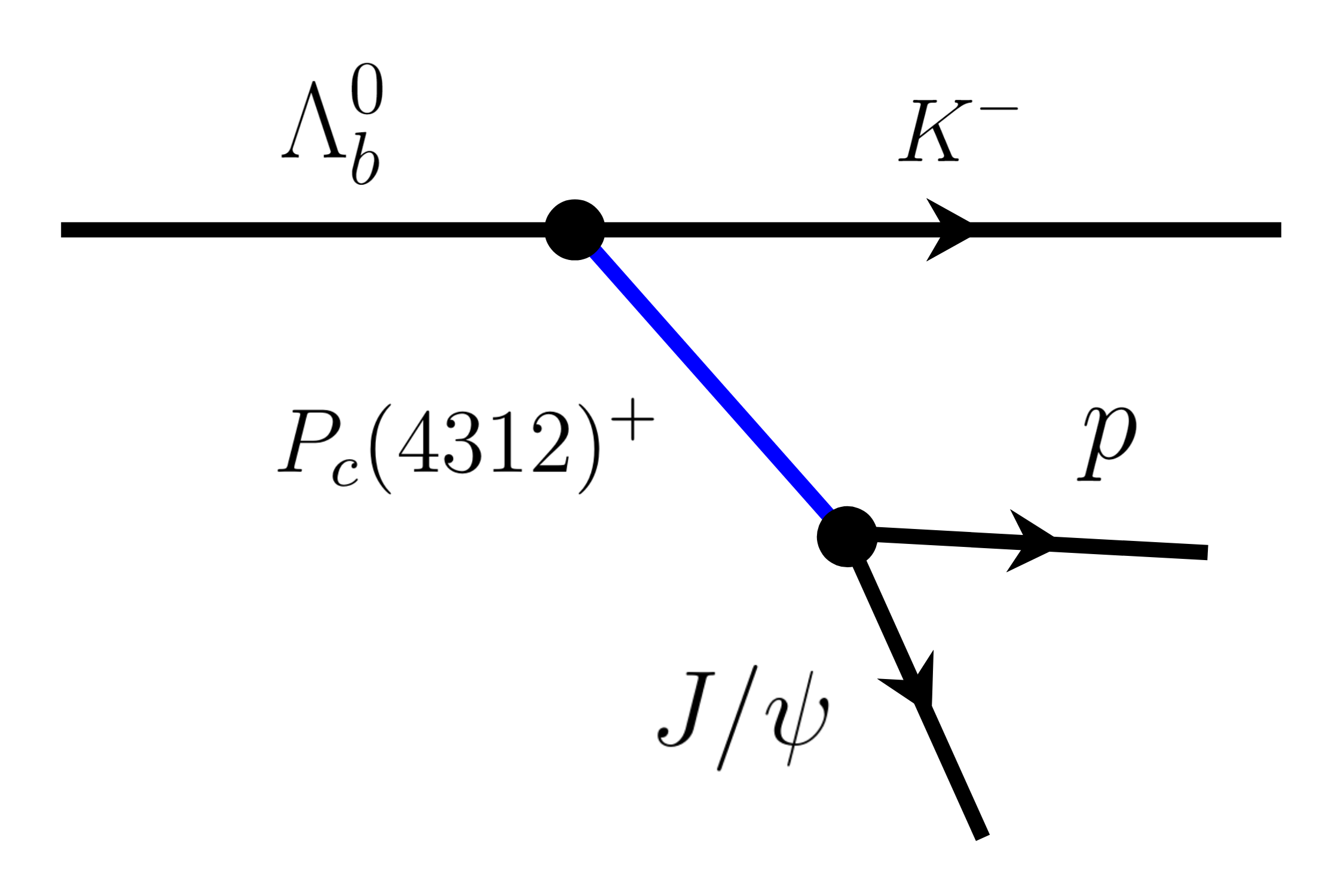}
		\caption[]{ }\label{fig:MechCompact}
		
	\end{subfigure}
	\begin{subfigure}[b]{0.47\columnwidth}   
		\centering 
		\includegraphics[scale = 0.115]{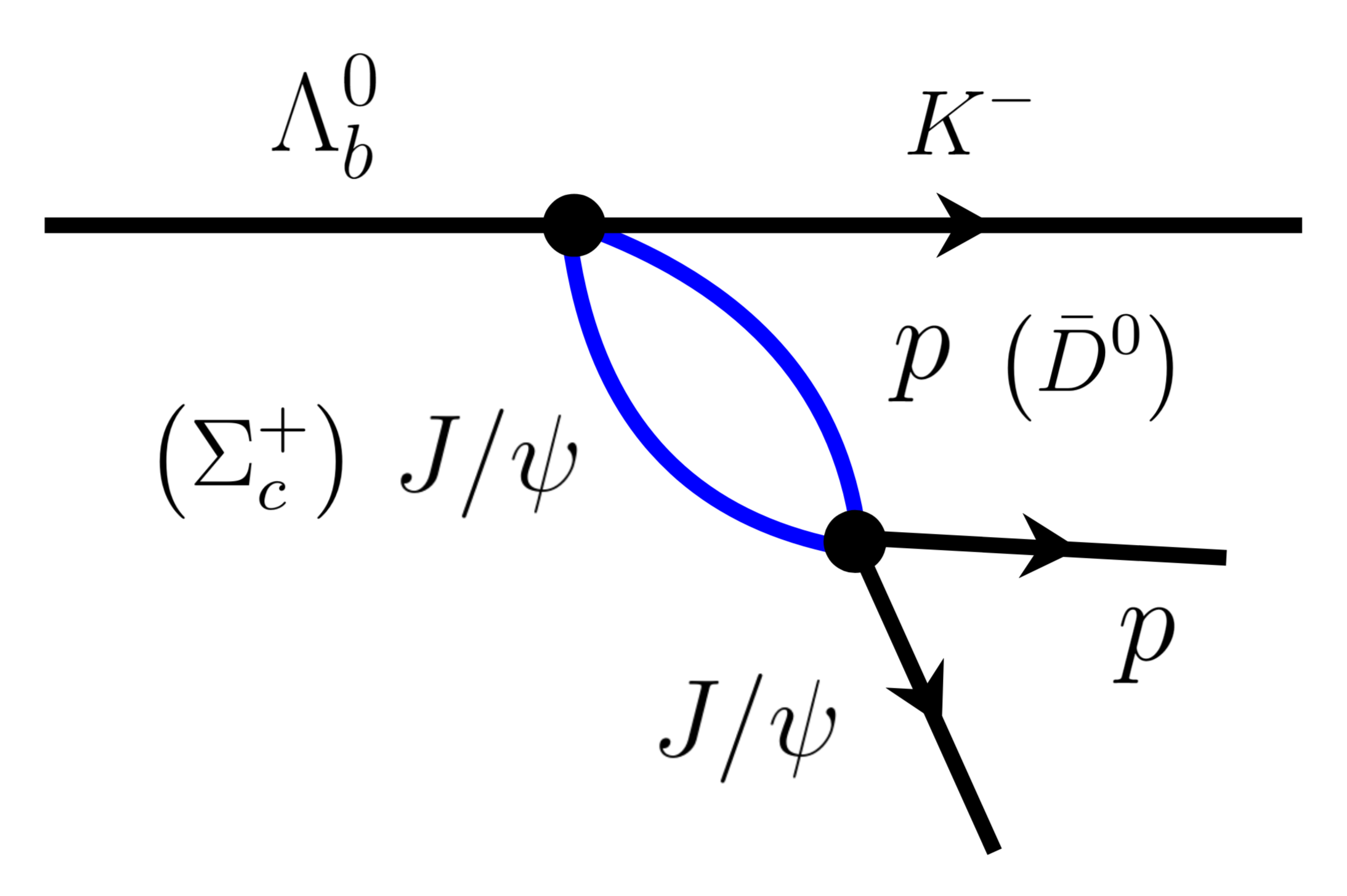}
		\caption[]{ }\label{fig:MechCoupled}
		
	\end{subfigure}
	\caption[]
	{Different mechanisms for the $\Lambda_b^0 \to J/\psi\,p K^-$ reaction that can produce a peak at 4312 MeV: a) triangle loop $\Lambda_c^+D_s^{**-}\bar{D}^{*0} $, b) triangle loop $J/\psi\Lambda^*p $, c) $P_\psi^N(4312)^+$ resonance, and d) coupled-channel decay.}
	\label{fig:mechs}
\end{figure*}

Furthermore, there are other triangle diagrams that can produce peaks in the $J/\psi p$ invariant mass \cite{Liu:2015fea,Bayar:2016ftu}. For instance, a $D_s^{**-}\Lambda_c^+\bar{D}^0$ triangle loop can generate an enhancement close to the $P_\psi^N(4312)^+$ state as well. In \cite{Bayar:2016ftu}, they considered multiple diagrams where a general $\Lambda^*$ hyperon is assigned as particle 1 (which can couple to $K^-p$), a $c\bar{c}$ meson for particle 2, and $p$ for particle 3. For each $c\bar{c}$ candidate, they used equations \eqref{eq:m23_range} and \eqref{eq:m1_range} to obtain the relevant range of the hyperon mass and the range of the resulting peak. Further analysis of the branching ratios led them to argue the plausibility of each diagram. Among these, if we select the $\Lambda^*J/\psi p$ loop with the mass of $\Lambda^*$ within [2151, 2523] MeV, a peak can be achieved within the energy region of [4035, 4366] MeV. This is relevant to the $P_\psi^N(4312)^+$ since the signal lies within the allowed energy region of the mechanism. 

Upon further inspection, the last two intermediate particles in this diagram, $J/\psi p$, are similar to the two final states in the mechanism. This sort of process must follow the Schmid theorem, which states that if particles 2 and 3 in the internal triangle diagram rescatters to similar final states 2+3, then the TS will simply be absorbed by the $s$-wave of a tree-level diagram and will not produce observable contributions \cite{Schmid:1967ojm}. In other words, this mechanism may also be ruled out if any manifestation of the triangle diagram is argued to not be detectable. However, it has been shown in \cite{Anisovich:1995ab} that if one considers an inelastic scattering process in the final vertex, the Schmid theorem becomes irrelevant as the triangle may still be present under coupled-channel effects. Moreover, if a physically plausible non-zero decay width $\epsilon$ is used in the triangle, the strength of its signal is found to be comparable to that of the tree-level diagram and will also no longer be negligible \cite{Debastiani:2018xoi}. 

In any case, we consider both aforementioned triangle diagrams in our study as we simply aim to simulate prominent line shapes from triangle mechanisms and investigate their contributions regardless of any physical constraints. These line shapes will be used to train the DNN. One may argue that doing this allows us to possibly ``trick" our machine and really test the limits of its abilities. It is up to the DNN to determine a more descriptive mechanism based solely on a pure line-shape analysis. Our benchmark is to assess if the DNN will consistently rule out these hypothetical triangle diagrams in favor of the dynamic pole structure for the $P_\psi^N(4312)^+$ state. Therefore, it follows that we take into account both these two triangle mechanisms in our study. From this point on, the first triangle diagram discussed will be referred to as triangle loop A, and the second as triangle loop B, as illustrated in Fig.~\ref{fig:TSloopA} and \ref{fig:TSloopB}, respectively.

\section{Uniformization of $S$-matrix}\label{sec:uniformization}
Enhancements observed in scattering cross-sections most often correspond to poles in the scattering matrix. Different $S$-matrix parametrizations can be used to analyze a given experimental data. Perhaps the most popular one is the $K$-matrix parametrization since it automatically satisfies the unitarity of the $S$-matrix. However, identifying the location and Riemann sheet of the poles requires solving a polynomial in channel momenta. As such, their locations cannot be controlled to ensure a well-distributed representation of different possible pole locations. Moreover, it is usually the case that a pole can be placed on the desired Riemann sheet in the $K$-matrix approach, but is often accompanied by shadow poles which may appear in the physical Riemann sheet. Such pole configuration will result in a violation of causality and must not be included in our training dataset. It is, therefore, more practical to use a parametrization in which poles can be controlled directly. It turns out that this problem can be addressed by a more rigorous approach called uniformization \cite{Santos:2023gfh}.

The idea of uniformization was introduced in \cite{Kato:1965,Newton:1982} to address the issue of branch points in a coupled two-channel problem. Recently, uniformization via the Mittag-Leffler parametrization of the amplitude was used to analyze newly discovered exotic hadrons \cite{Yamada:2020rpd,Yamada:2021cjo,Yamada:2021azg}. To construct a uniformized $S$-matrix, we first define some kinematical variables. The on-shell mass condition of a two-particle collision is given by
\begin{equation}
	\sqrt{s} = \sqrt{p^2 + m_{1}^2} + \sqrt{p^2 + m_{2}^2},
\end{equation}
where $\sqrt{s}$ is the scattering energy, $p$ is the breakup momentum, and $m_1$, $m_2$ are the masses of the particles. From this, we can show that
\begin{equation}
	p = \frac{\sqrt{s - \epsilon^2}\sqrt{s - \epsilon(\epsilon - 4\mu)}}{2\sqrt{s}},
\end{equation}
where $\epsilon=m_1 + m_2$ is the threshold, and $\mu$ is the reduced mass. Notice that even though we write $p$ in terms of $\sqrt{s}$, there are still branch point singularities due to the existence of square root cross terms. To resolve this, we introduce the uniformized momentum $q_i$ to simplify the scattering energy, giving us
\begin{equation}
	s = q^2 + \epsilon^2,
\end{equation}
and
\begin{equation}
	s = q_1^2 + \epsilon_1^2 = q_2^2 + \epsilon_2^2,
	\label{UniMont}
\end{equation}
for a two-channel scattering.

The two-channel scattering can be described by a $2\times 2$ uniformized $S$-matrix with elements constructed using a Jost function $D$ and given by
\begin{equation}
	\begin{aligned}
		S_{11}(q_1,q_2)&=\dfrac{D(-q_1,q_2)}{D(q_1, q_2)},\\
		S_{22}(q_1,q_2)&=\dfrac{D(q_1,-q_2)}{D(q_1, q_2)}
	\end{aligned}
\end{equation}
and
\begin{equation}
	S_{12}^2=S_{11}S_{22}-
	\text{det}S.
\end{equation}
Instead of expressing the Jost function in terms of the uniformized momentum $q$, we define a uniformized variable $\omega$ given by
\begin{equation}
	\omega=\frac{q_1 + q_2}{\sqrt{\epsilon_2^2 - \epsilon_1^2}}; \quad \frac{1}{\omega} = \frac{q_1 - q_2}{\sqrt{\epsilon_2^2 - \epsilon_1^2}}
\end{equation}
which can be derived by utilizing the two expressions in equation \eqref{UniMont}. We can then construct a Jost function with the form
\begin{equation}
	D(\omega) \propto \frac{1}{\omega^2}(\omega - \omega_\text{pole})(\omega + \omega_\text{pole}^*)(\omega - \omega_\text{reg})(\omega + \omega_\text{reg}^*).
\end{equation}
The factor $(\omega - \omega_\text{pole})$ produces the poles in the $S$-matrix. The $1/\omega^2$ prohibits very energetic initial states from having no final states after the interaction as this is unphysical. The pole $\omega_\text{reg}$ ensures that the diagonal elements go to unity ($S_{ii} \rightarrow 1$) when the initial states have very large momentum ($\omega \rightarrow \infty$), which is a well-known property in potential scattering theory \cite{Taylor}. It is controlled such that it is far from the relevant pole and does not affect the interpretation of the enhancements. To satisfy hermiticity below the lowest threshold, there must be a pole in the negative complex conjugate of $\omega_\text{pole}$, that is why we have the factors ($\omega + \omega_\text{pole}^*$)($\omega + \omega_\text{reg}^*$).

\begin{figure}[h!!!]
	\centering
	\includegraphics[width=0.9\linewidth]{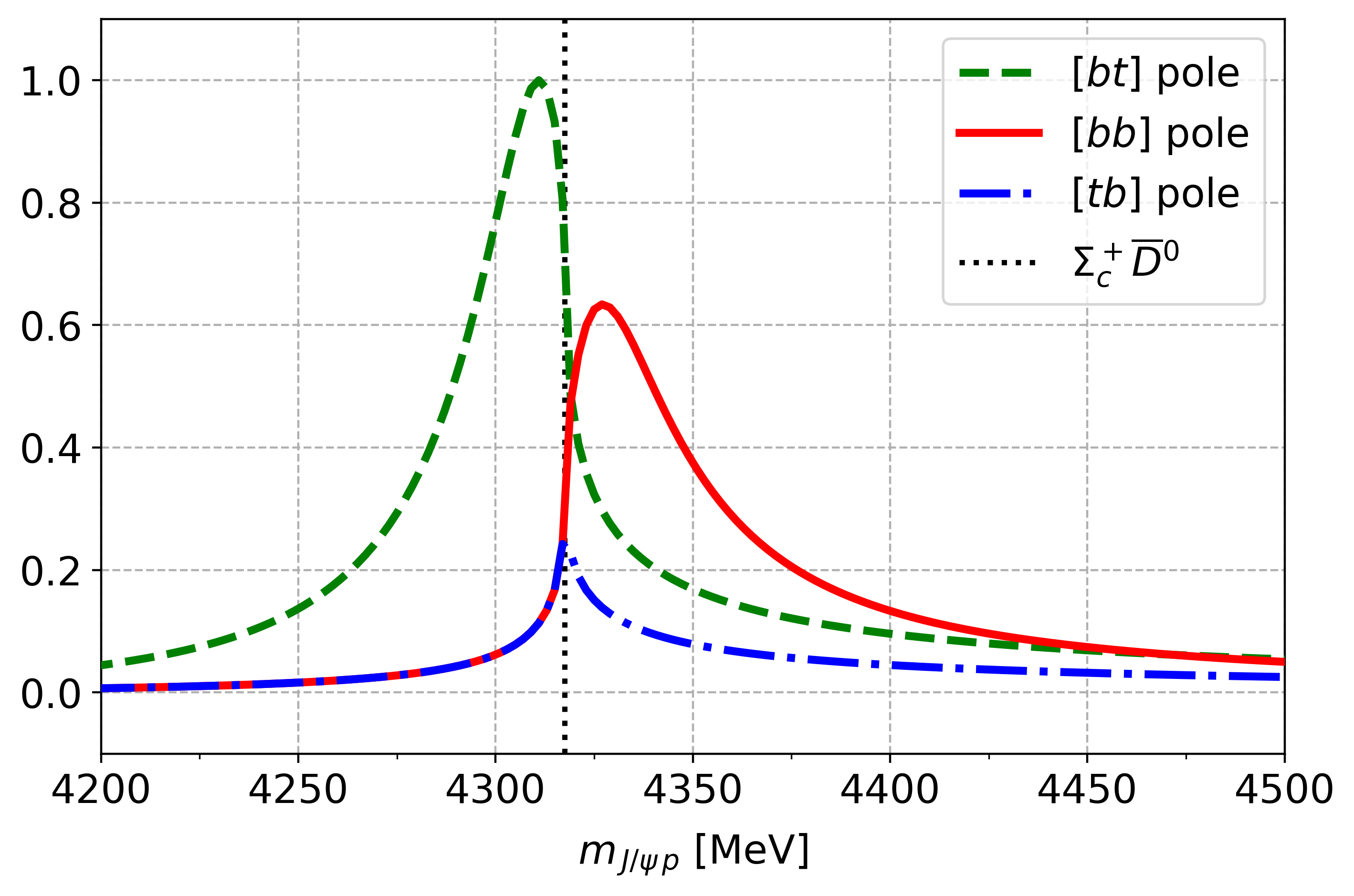}
	\caption[]{Expected line shapes arising from isolated poles in one of the unphysical Riemann sheets. The black vertical dotted line is the second threshold. The energy poles were generated with real part exactly at the second threshold and imaginary part of $-10$ MeV.}\label{fig:btbbtbpeaks}
\end{figure}

Hence, by using uniformization, one has the freedom to adjust the position of the pole and its regulator to their desired Riemann sheets while preserving causality. In two-channel scattering, there are four Riemann sheets of the complex energy plane. The first Riemann sheet named [$tt$]-sheet is called the physical sheet where the scattering region is located. There should be no poles in the physical sheet to obey causality \cite{Kampen:1953_1,Kampen:1953_2}. The other three Riemann sheets, namely [$bt$], [$bb$], and [$tb$]-sheets, are the unphysical sheets where poles located can be associated with the presence of an unstable quantum state in the reaction involved. One can say that the pole is located in the [$bt$]-sheet when the peak is below the second threshold, and a pole is located in the [$tb$]-sheet when the peak is located exactly at the threshold. These pole structures, and their manifestation on the physical region shown in Fig.~\ref{fig:btbbtbpeaks}, are useful in the description of the hadron's internal structure as demonstrated in the pole-counting argument by Morgan and Pennington \cite{Morgan:1992,MorganPennington:1991,MorganPennington:1993}. The location of the Riemann sheets is shown in Table \ref{tab:RiemannSheets}. The pole regulator in $D(\omega)$ has the form $\omega_\text{reg} = \exp(-i\pi/2)/ |\omega_\text{pole}|$ where the phase factor forces the regulator to be in either the farther [$bb$]-sheet or the [$tb$]-sheet below the lowest threshold.

\begin{table}[h!]
	\centering
	{\renewcommand{\arraystretch}{1.5}
		\begin{tabular}{c@{\hskip 0.1in}c@{\hskip 0.1in}c}
			\hline\hline
			Sheet & Topology in complex $E$ & Range of Values \\   
			\hline
			$[tt]$ & $\theta_1 \in (0,2\pi)$; $\theta_2 \in (0,2\pi)$ & $\text{Im}q_1 > 0$ and $\text{Im}q_2 > 0$ \\
			\hline
			$[bt]$ & $\theta_1 \in (2\pi, 4\pi)$; $\theta_2 \in (0,2\pi)$ & $\text{Im}q_1 < 0$ and $\text{Im}q_2 > 0$ \\
			\hline
			$[bb]$ & $\theta_1 \in (2\pi,4\pi)$; $\theta_2 \in (2\pi,4\pi)$ & $\text{Im}q_1 < 0$ and $\text{Im}q_2 < 0$ \\
			\hline
			$[tb]$ & $\theta_1 \in (0,2\pi)$; $\theta_2 \in (2\pi,4\pi)$ & $\text{Im}q_1 > 0$ and $\text{Im}q_2 < 0$ \\
			\hline\hline
	\end{tabular}}
	\caption{Riemann sheet notation and classification. We follow Pearce and Gibson's sheet-naming convention \cite{PearceGibson}.}
	\label{tab:RiemannSheets}
\end{table}

The $J/\psi p$ scattering is a coupled-channel scattering problem where the first channel is set to always be $J/\psi p$ and the second channel is $\Sigma_c^+\Bar{D}^0$ as this is the prominent threshold near the $P_\psi^N(4312)^+$ signal. One can construct its uniformized $S$-matrix where the elements $S_{11}$ describes the transition of $J/\psi p \rightarrow J/\psi p$, $S_{12}$ for $J/\psi p \rightarrow \Sigma_c^+ \Bar{D}^0$, $S_{21}$ for $\Sigma_c^+ \Bar{D}^0 \rightarrow J/\psi p$, and $S_{22}$ for $\Sigma_c^+ \Bar{D}^0 \rightarrow \Sigma_c^+ \Bar{D}^0$. A compact state arises when there are poles both in the [$bt$] and [$bb$]-sheets, corresponding to the mechanism in Fig.~\ref{fig:MechCompact}. By placing two identical energy poles on these sheets (i.e. the poles are equidistant from the relevant threshold), one can interpret the fitted enhancement as a pure compact state similar to fitting a Breit-Wigner amplitude \cite{Baru:2003qq}. On the other hand, a quasi-bound state interpretation is possible whenever there is an isolated pole in the $[bt]$-sheet with strong coupling to the higher channel, corresponding to Fig.~\ref{fig:MechCoupled}. However, this interpretation should be examined carefully since a virtual state of the higher channel can move to the second Riemann sheet due to strong channel-coupling \cite{Fernandez-Ramirez:2019koa, Frazer:1964, PearceGibson}. It must be pointed out that bound and virtual states fall under the same general classification of hadronic molecules as demonstrated in \cite{Matuschek:2020gqe}. In any case, any interpretation should be supported by the corresponding pole trajectory, which can only be done in a model-dependent way \cite{Badalyan:1982,Baru:2003qq, Hanhart:2014ssa}. 

\begin{figure}[h!!!]
	\centering
	\includegraphics[width=\linewidth]{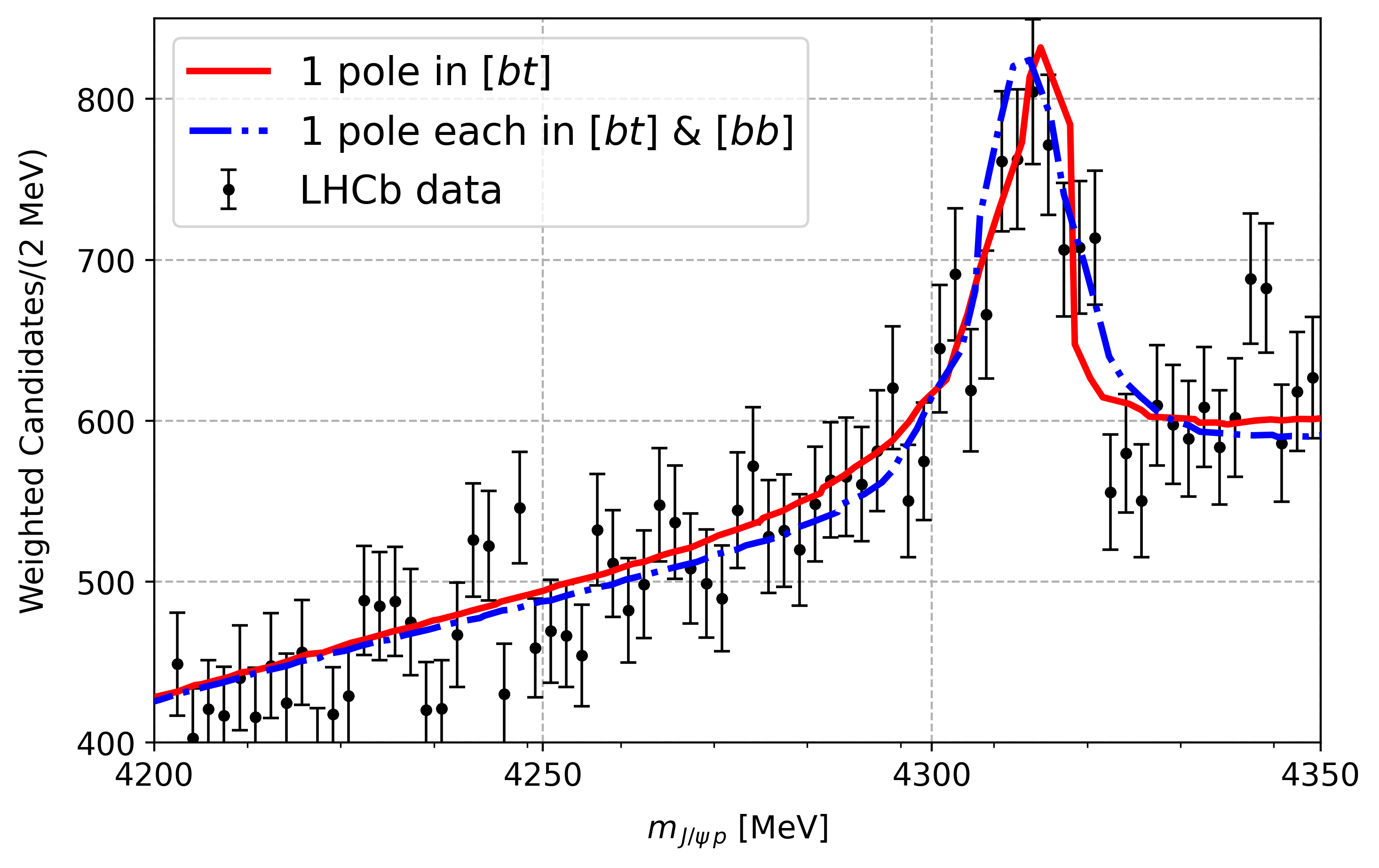}
	\caption[]{The $\cos\theta_{P_c}$-weighted $J/\psi p$ invariant mass distribution with fitted uniformized $S$-matrix amplitudes: 1 pole in [$bt$] (red), and 1 pole each in [$bt$] and [$bb$] (blue).}\label{fig:Pc4312_btbb}
\end{figure}

Through the uniformization scheme, one can easily construct pole-based amplitudes that correspond to the interpretations discussed. This is demonstrated in Fig.~\ref{fig:Pc4312_btbb}. We assign an isolated pole with energy (in units of MeV) at $4317.49 - 7j$ in the [$bt$]-sheet to construct an amplitude representing a possible hadronic molecule, while we assign one pole each at $4311.90 - 7j$ in both the [$bt$] and [$bb$]-sheets representing a compact pentaquark state. An amplitude is generally complex-valued with a real and imaginary part, but for our purposes, it is sufficient to consider its modulus square which corresponds to the intensity. Therefore, plotting the $|T_{11}|^2$ ($T$-matrix) contribution, where $S=1-2iT$, using the uniformized poles and following equation \eqref{eq:full_amplitude}, we see that the two similar line shapes fit the LHCb experimental data, both with a prominent peak at 4312 MeV. If we compare our demonstrations in Fig.~\ref{fig:Pc4312_btbb} and \ref{fig:Pc4312_TSBW}, we notice that the pole-based enhancements have a seemingly narrower and asymmetric width compared to that produced by a triangle singularity, and can fit the experimental error bars better especially near the $\Sigma_c^+\Bar{D}^0$ threshold. Such asymmetry in the line shape is one of the characteristic features whenever the hadrons in the higher mass channel form a hadronic molecule. We take note of these observations as we move on to constructing the training dataset for our DNN.

\section{Construction of Deep Neural Network}\label{sec:DNN}
\subsection{Generation of training datasets}

\begin{figure}[ht!]
	\centering
	\includegraphics[width=0.95\linewidth]{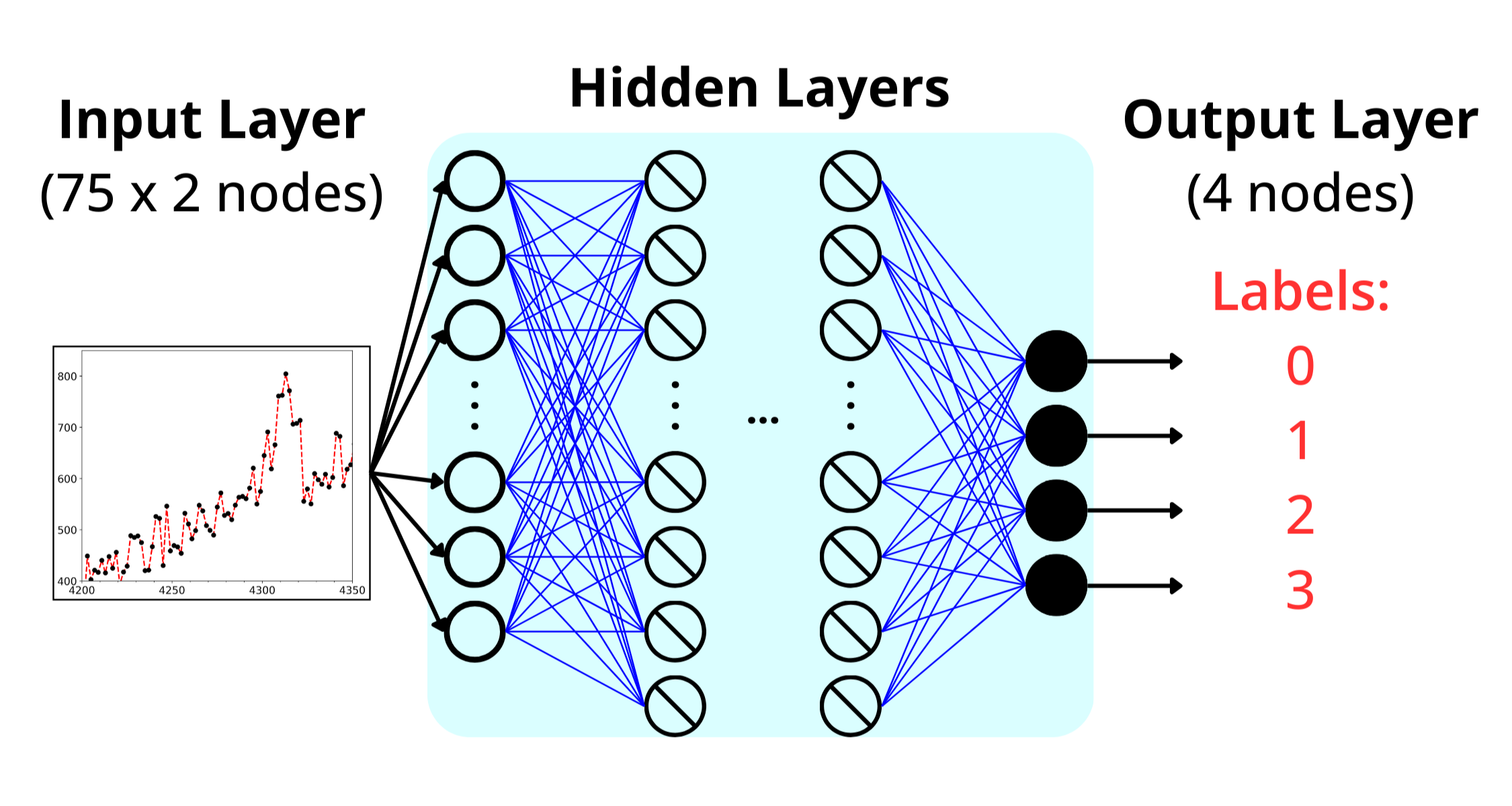}
	\caption[]{DNN architecture. Circles are for input nodes, cross circles are for hidden layer nodes, and shaded circles are for output nodes. The blue lines refer to the weights and biases.}\label{fig:DNN}
\end{figure}

The experimental data is an invariant mass distribution containing a set of points specified by the energy value (horizontal axis) and the intensities (weighted candidates in the vertical axis). It is important to consider the energy values as part of our input features since they are expected to be smeared out (not equidistant) due to resolution effects of the particle detector. Thus, our goal is to construct a DNN model that accepts two-dimensional line shapes as input features and maps them to output targets corresponding to the possible classifications. Since we will eventually apply the trained DNN on interpreting the experimental data, the training dataset must follow the general characteristics of the experimental data (similar to our demonstrations in Fig.~\ref{fig:Pc4312_TSBW} and \ref{fig:Pc4312_btbb}). In the LHCb data \cite{hepdata.89271}, the energy axis is divided into bins of 2 MeV each. For our training dataset, we can generate amplitudes on randomly spaced energy points, where one representative energy value is randomly chosen for each bin. We use a uniform distribution for this task. Moreover, we restrict our region of interest to the energy region of 4200 MeV to 4350 MeV, in order to avoid other prominent structures aside from the $P_\psi^N(4312)^+$. This region consists of 75 energy bins, implying that our line shapes shall have 75 points each for the energy value and weighted candidates. This totals to 150 input nodes, as presented in Fig.~\ref{fig:DNN}.

For the output targets, we create labels for each of the possible classifications, as listed in Table~\ref{tab:labels}. The triangle singularity takes up the first label, while the rest are possible pole configurations. As previously stated, 1 pole in [$bt$] may be associated to the hadronic molecule interpretation, 1 pole in [$tb$] corresponds to a virtual state, and 1 pole in [$bt$] and 1 pole in [$bb$] can accommodate a compact pentaquark interpretation. Refer to section~\ref{sec:uniformization} for the detailed discussion. It is possible to include more sophisticated pole configurations, such as considering 2 or more poles in each of the sheets as done in \cite{Sombillo:2021rxv}. However, for the purposes of our task which is mainly to distinguish triangle singularity from the dynamic pole structure, the simple pole configurations shall suffice. Our next step is to then generate an equal number of training datasets for each of these classifications.

As previously discussed, there are multiple triangle diagrams that can produce enhancements within our energy region. For our task, we only consider the two loops, shown in Fig.~\ref{fig:TSloopA} and \ref{fig:TSloopB}, that can produce a peak exactly at 4312 MeV so that our trained DNN can later match it to the experimental data. Inclusion of other relevant loops may be done to expand the training dataset but is not mandatory. We highlight once again that it is sufficient to use these possibly unphysical diagrams, as we aim to test the likelihood of these mechanisms through the DNN. Moreover, we have shown that by varying the masses of the intermediate particles, one can construct enhancements at different locations. For this, we can consider the uncertainty in the masses of the hadrons involved, as well as vary the mass of particle 1, using equation \eqref{eq:m1_range}, such that the produced enhancement still falls within our region of interest of [4200,4350] MeV. For triangle loop B, varying particle 1 means considering the different $\Lambda^*$ hyperon candidates as listed in PDG. For triangle loop A, it is simply adjusting the mass of the hypothetical $D_s^{**-}$ meson. It is also possible to vary the $\Lambda$ cut-off parameter and the $\epsilon$ decay width to control the strength and distribution of the resulting peaks. 

Furthermore, it is important to recognize the possibility of producing cusp around the relevant threshold in the triangle singularity. In principle, one cannot separate the possible appearance of cusp in the triangle mechanism due to the presence of two-hadron interaction leading to the final state. In order to ensure that the triangle singularity is the dominant structure in the line shape, we ensure that the cusp effect is small by making the cut-off parameter very large. This means that the poles produced by the effective meson-baryon interaction is already far from the relevant scattering region. The cusp at the relevant threshold will still be present but it is negligible in comparison to the triangle enhancement.

All the range of values used for these parameters are listed in Table \ref{tab:parameters}. By randomly choosing a certain number of values for each of the parameters and taking all their combinations, we construct 5000 line shapes each for triangle loop A and triangle loop B. For convenience, we call these datasets T00A and T00B, respectively. Note that to mimic the experimental data, we include a projected phase space and an arbitrary background polynomial (noise) to the amplitudes following equation \eqref{eq:full_amplitude}. A total of 10,000 training data is thus constructed for the triangle singularity.

\begin{table}[]
	\centering
	{\renewcommand{\arraystretch}{1.5}
		\begin{tabular}{lc}
			\hline\hline
			Parameter\;        & Range of values [MeV] \\ \hline
			$m_{\Lambda_b^0}$  & $5619.60 \pm 0.17$  \\
			$m_{K^-}$          & $493.677 \pm 0.016$ \\
			$m_{\Lambda_c^+}$  & $2286.46 \pm 0.14$  \\
			$m_{\bar{D}^{*0}}$ & $2006.85 \pm 0.05$  \\
			$m_{J/\psi}$       & $3096.9 \pm 0.006$  \\
			$m_{D_{s}^{**}}$   & [3209.80, 3315.00]  \\
			$m_{\Lambda^*}$    & [2490.00, 2522.70]  \\
			$\Lambda$          & [2000.0, 2500.0]     \\
			$\epsilon$         & [1.0, 10.0]          \\ \hline\hline
	\end{tabular}}
	\caption{Range of values of each parameter used in the generation of training dataset for triangle singularity line shapes. The first 5 listed are hadron masses retrieved from PDG \cite{Workman:2022ynf}. The next two are ranges derived from equation \eqref{eq:m1_range}, and the last two are arbitrarily defined.}\label{tab:parameters}
\end{table}

For the pole-based enhancements, we follow the formalism discussed in section~\ref{sec:uniformization}. The parameter that we vary here is the location of the poles to be placed in the various Reimann sheets. We allow our poles to be randomly placed such that
\begin{equation*}
	\begin{cases}
		T_2-50 \leq \text{Re}\,E_\text{pole} \leq 4350\; & \text{all RS}\\
		-100 \leq \text{Im}\,E_\text{pole} < 0 & \text{[$bt$] \& [$bb$]}\\
		0 < \text{Im}\,E_\text{pole} \leq 100 & \text{[$tb$]}
	\end{cases}
\end{equation*}
where energy $E_\text{pole}$ is in MeV and $T_2$ is the $\Sigma_c^+\Bar{D}^0$ threshold energy. In addition to the main poles, we also produce distant background poles far from the scattering region, in order to simulate the smooth non-resonant background phase. Similarly, we include the projected phase space and arbitrary background polynomial to the amplitudes. By randomly generating 100 $\text{Re}\,E_\text{pole}$ and 100 $\text{Im}\,E_\text{pole}$ and taking their combinations, we construct 10,000 line shapes for each of the pole configurations listed in Table~\ref{tab:labels}. We name these datasets as P01, P02, and P03. A sample and short discussion of all these training datasets can be found in appendix \ref{sec:appendix}. Finally, we compile all our data with their corresponding proper output labels and prepare for DNN training.

\begin{table}[h!]
	\centering
	{\renewcommand{\arraystretch}{1.5}
		\begin{tabular}{lcc}
			\hline\hline
			Output Label       & Model & Input Dataset \\ \hline
			\multirow{2}{*}{0} & Triangle Loop A & T00A \\
			& Triangle Loop B & T00B \\
			1 & 1 pole in $[bt]$ & P01 \\
			2 & 1 pole in $[tb]$ & P02 \\
			3 & 1 pole each in $[bt]$ and $[bb]$ & P03 \\ \hline\hline
	\end{tabular}}
	\caption{Classification output labels and corresponding input datasets. Note that T00A and T00B datasets are generated using different triangle diagrams, but must share the same triangle singularity classification under output label (0).}\label{tab:labels}
\end{table}

\subsection{DNN architecture}

The basic architecture of our DNN is shown in Fig.~\ref{fig:DNN}. The input layer has 75 nodes for the energy points and 75 nodes for the intensities. In the output layer, we create 4 nodes such that each corresponds to a classification or label listed in Table~\ref{tab:labels}. In this study, we consider three different architectures enumerated in Table~\ref{tab:dnnmodels}. The notation [XXX-...-XXX] denotes the number of hidden layers and the number of nodes in each hidden layer. For example, [250-100] refers to an architecture with 250 nodes in the first hidden layer and 100 nodes in the second hidden layer. For a complete notation, we can also write 150-[250-100]-4 to include the input and output layers. For the hidden layers, we utilize the different DNN architectures used in \cite{Sombillo:2021ifs} which are already tested and validated for the bound-virtual classification of single-channel scattering. For a more rigorous neural network architecture search, one may utilize the Bayesian optimization \cite{zoph2017neural,snoek2012practical,frazier2018tutorial}. In the present study, the architectures shown in Table~\ref{tab:dnnmodels} should suffice for our purpose.

The basic operation of a typical feed-forward neural network is described as follows: A linear combination is formed using the node values of the preceding layer and the present weight and biases of the network. Except for the input layer, the hidden and the output layers are further equipped with activation functions which makes the mapping non-linear. 
We use the rectified linear unit (ReLU) as the activation function, and the softmax for the output layer. As is standard in a classification problem, the softmax cross entropy is used as the cost function to be minimized \cite{Aggarwal:2018}. 
This combination of linear transformations and non-linear effects of the activation functions allows the DNN to map the input space to the output space, which are further improved by adjusting the weights and biases during the training iteration. We follow the Chainer framework \cite{Chainer:2015,Chainer:2017,Chainer:2019} in executing the construction and training of our DNN models. All the codes used in this study are accessible through our public repository in \cite{MyGithub}.

\begin{table}[]
	{\renewcommand{\arraystretch}{1.5}
		\begin{tabular}{c@{\hskip 0.2in}l}
			\hline\hline
			\textbf{DNN model}& \textbf{Optimizer and architecture} \\
			\hline
			DNN 1& AdaGrad: 150-[250-100]-4  \\
			DNN 2& AdaGrad: 150-[250-100-50]-4 \\
			DNN 3& AdaGrad: 150-[250-250-250]-4 \\
			\hline
			DNN 4& AMSGrad: 150-[250-100]-4  \\
			DNN 5& AMSGrad: 150-[250-100-50]-4 \\
			DNN 6& AMSGrad: 150-[250-250-250]-4 \\
			\hline
			DNN 7& SMORMS3: 150-[250-100]-4  \\
			DNN 8& SMORMS3: 150-[250-100-50]-4 \\
			DNN 9& SMORMS3: 150-[250-250-250]-4 \\
			\hline\hline
	\end{tabular}}
	\caption{DNN models used in the final analysis.}
	\label{tab:dnnmodels}
\end{table}

\subsection{Training and testing}
\begin{figure*}[ht!]
	\centering
	\includegraphics[scale=0.4]{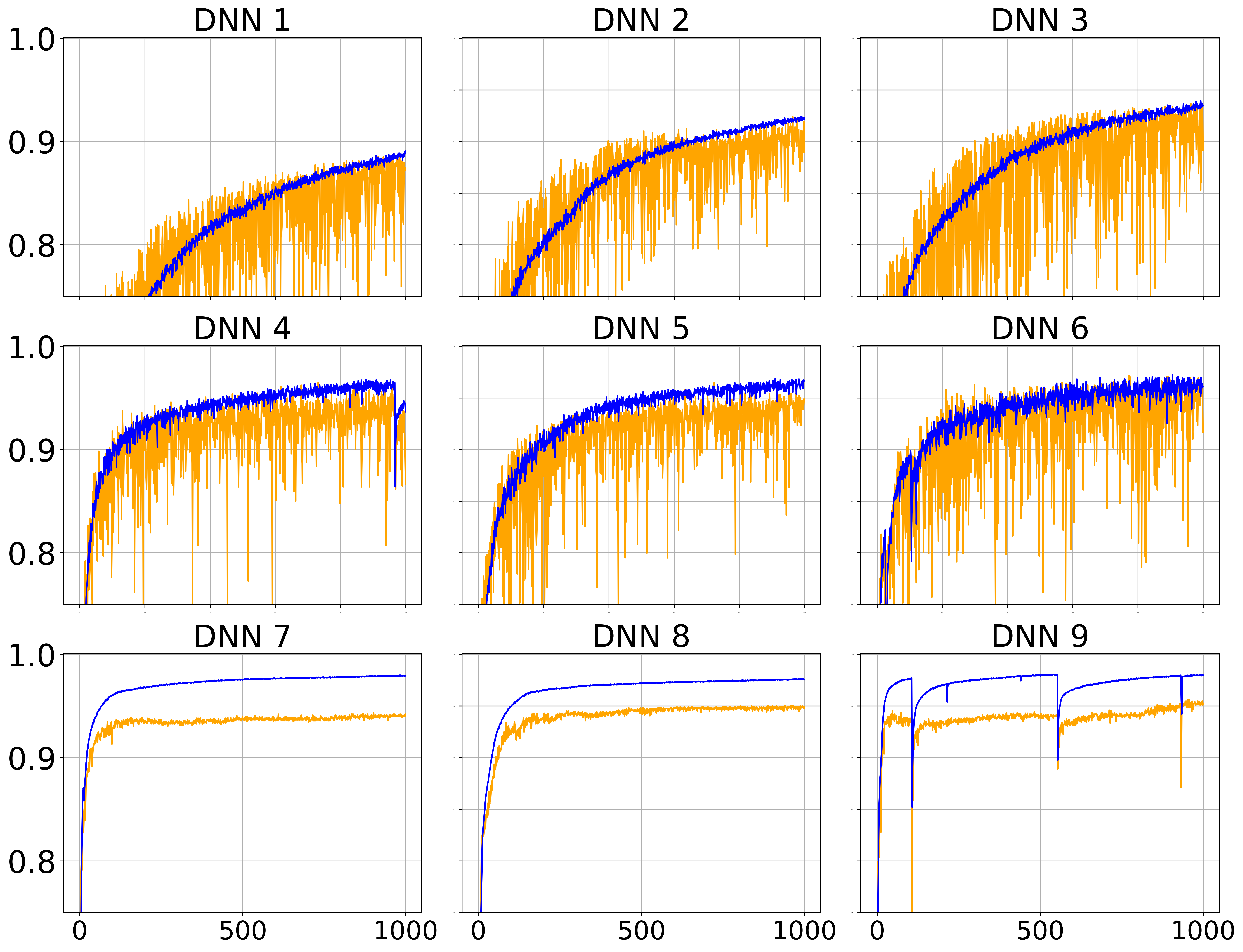}
	\caption[]{Training performance of DNN models in Table~\ref{tab:dnnmodels}. The blue lines are for the training dataset and the orange lines are for the testing set.}
	\label{fig:training_ada}
\end{figure*}

During training, the prepared dataset is fed into the DNN using some mini-batch procedure to add stochasticity in
estimating the cost-function. It must be noted that we independently produced a separate set of ``testing" data that is hidden from the DNN during training. The testing dataset consists of 320 line shapes for each classification, constructed using completely different sets of parameter values from the training dataset. This is done to ensure that the DNN is not simply ``memorizing" and to check if the model can generalize beyond the training set. The testing data are only fed into the DNN after every training epoch to measure the accuracy of the trained model. During every loop, the weights and biases of the nodes are being optimized following some framework for gradient descent. For this, we use different optimizers available in Chainer \cite{Chainer:Optimizers} and mix them with our chosen architectures. Different optimizers are used such as Adam, AdaDelta, AdaGrad,AMSGrad, AdaBound, AMSBound, RMSprop, RMSpropGraves, and SMORMS3. We find that some models learn the classification problem but are eventually reduced to random models after a few more training epochs. There are also models that do not learn the problem at all. This is evidence that disentangling the nature of enhancements is a non-trivial task. For the different combinations of chosen architecture, only the AdaGrad, AMSGrad, and SMORMS3 show promising results. We label our chosen models according to  Table~\ref{tab:dnnmodels} for reference.

Fig.~\ref{fig:training_ada} illustrates the various DNN models capable of learning the classification problem. The SMORMS3 (squared mean over root mean squared cubed) optimizer demonstrates rapid convergence, achieving higher training and testing accuracy compared to AdaGrad and AMSGrad. SMORMS3 is particularly effective in handling gradients with high variance, making it suitable for DNNs with multiple hidden layers \cite{SMORMS3_1,SMORMS3_2}. While SMORMS3 shows a divergence between testing and training results, the primary focus remains on achieving high accuracy, which SMORMS3 consistently delivers. Additionally, other optimizers such as AdaGrad and AMSGrad also achieved high accuracy, further emphasizing the robustness of these methods.

Estimating the uncertainty of the reported accuracy, ideally without additional computational cost during training, is crucial. This can be achieved using the snapshot ensemble method, which builds an ensemble of models from different epoch states of the DNN \cite{huang2017snapshot}. Following this principle, we trained nine models for up to 1,050 epochs. We then used the accuracy values from epochs 950 to 1,050 to calculate the mean and standard deviation. The standard deviation of the 100 epoch states of a particular DNN model provides our estimation of the uncertainty in the accuracy. The results are presented in Table~\ref{tab:traintest_sd}. In addition to the estimation of uncertainty in the accuracy, the snapshot ensemble method will be used in the final inference stage.
\begin{table}[ht!]
	\centering
	\renewcommand{\arraystretch}{1.5}{
		\begin{tabular}{>{\bfseries}lcccc}
			\hline\hline
			\multicolumn{1}{c}{\textbf{DNN Model}} & \multicolumn{2}{c}{{}\quad\textbf{Training}} & \multicolumn{2}{c}
			{{}\quad\textbf{Testing}} \\
			\cline{2-5}
			&{}\quad\textbf{Mean} & \textbf{SD} & {}\quad\textbf{Mean} & \textbf{SD} \\
			\hline\hline
			DNN 1 
			&0.829 & 0.056 
			&0.811 & 0.062 \\
			DNN 2
			&0.876 & 0.046 
			&0.863 & 0.052 \\
			DNN 3
			&0.890 & 0.044 
			&0.866 & 0.063 \\
			\hline\hline
			DNN 4
			&0.945 & 0.016 
			&0.920 & 0.028 \\
			DNN 5
			&0.943 & 0.023 
			&0.915 & 0.036 \\
			DNN 6
			&0.944 & 0.021 
			&0.922 & 0.040 \\
			\hline\hline
			DNN 7
			&0.975 & 0.004 
			&0.937 & 0.003 \\
			DNN 8
			&0.971 & 0.005 
			&0.944 & 0.005 \\
			DNN 9
			&0.974 & 0.002 
			&0.940 & 0.013 \\
			\hline\hline
		\end{tabular}
		\caption{Training and Testing Accuracy with estimated mean and uncertainty (SD) using the snapshot ensemble method for each DNN model listed in Table~\ref{tab:dnnmodels}.}
		\label{tab:traintest_sd}}
\end{table}

\section{Application of Trained Deep Neural Network}\label{sec:results}
\subsection{Validation of trained DNN models}

\begin{figure*}[ht!]
	\centering
	\includegraphics[scale=0.3]{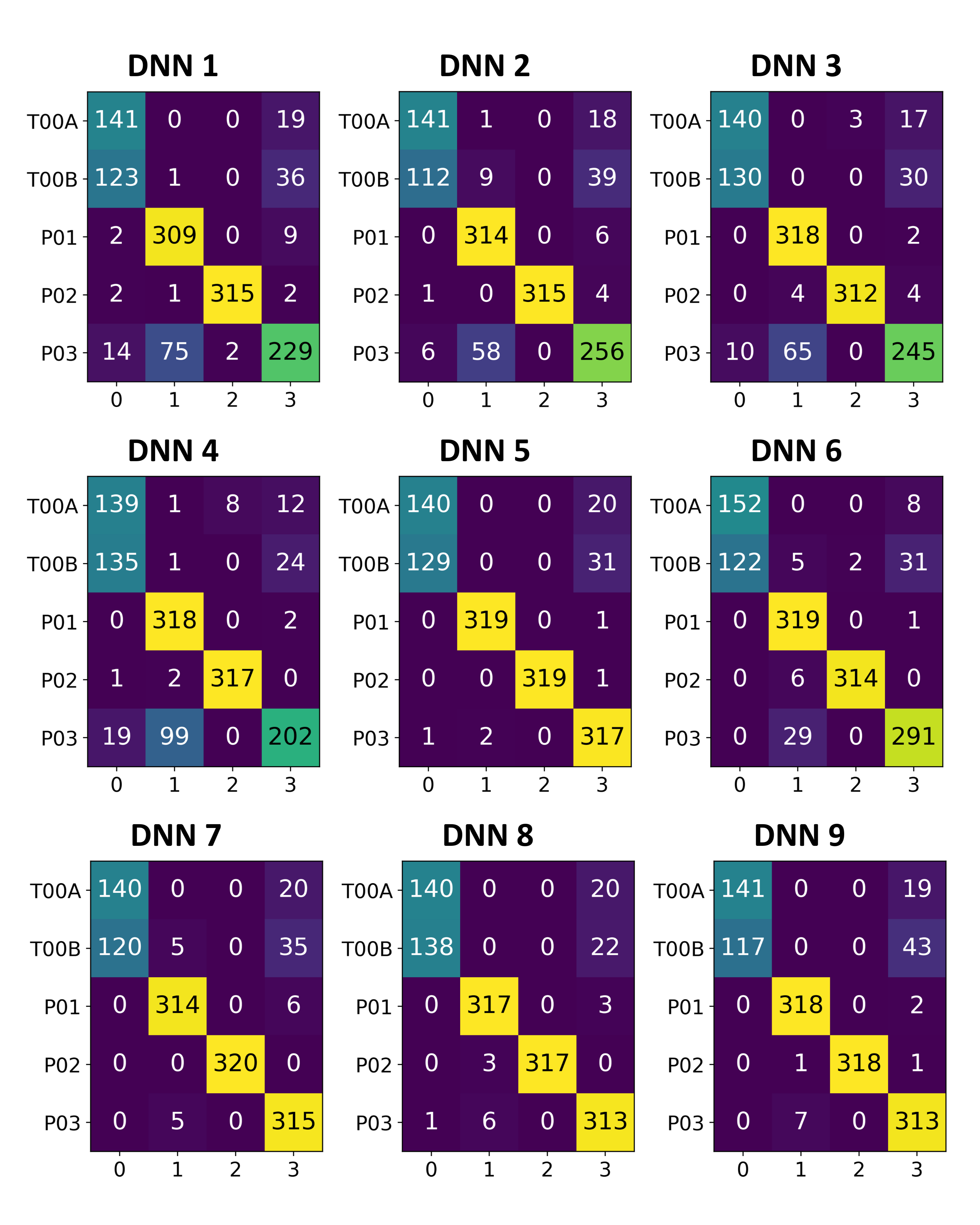}
	\caption[]
	{Confusion matrices for the nine DNN models. T00A, T00B, P01, P02, and P03 are the true labels while 0,1,2, and 3 are the predicted labels. The architectures and optimizers used are shown in Table~\ref{tab:dnnmodels}.}
	\label{fig:confusion_matrix}
\end{figure*}

During training, the testing accuracy plateaued below the training accuracy, indicating that some classifications in our training dataset are difficult to distinguish. To identify these confusing line shapes, we calculated the confusion matrix, focusing on the state of each DNN at epoch 1000. For validation, we generated an independent set of 320 datasets per classification. It is also useful to separate the two triangle enhancements considered in the study to determine which mechanism is most likely confused with the pole-based structure. Note that both T00A and T00B datasets produced from triangle mechanisms must fall under the label (0) to be considered correct. Fig.~\ref{fig:confusion_matrix} shows the confusion matrices for the different DNN models, with rows corresponding to the true labels and columns to the predictions. Notice that the triangle mechanism in Fig.~\ref{fig:TSloopA} (label T00A) is less likely to be confused with the pole-shadow pole configuration (label 3) compared to the other triangle. One possible reason for this distinction is the presence of cusp around the $\Lambda_c^+\bar{D}^{*0}$ threshold due to the final state interaction, despite being small, might be aiding the DNN in the classification. On the other hand, the second triangle diagram has no such cusp structure within the energy region considered in the analysis since $J/\psi p$ is already the lowest mass channel. Consequently, the triangle mechanism in Fig.~\ref{fig:TSloopB} (label T00B) will most likely resemble a Breit-Wigner like line shape (label 3) as is evident in the confusion matrices. If the final inference using the experimental data favor either the label (0) or (3), then further analysis must be done.

In general, The DNN models perform well with classifying the pole-based enhancements, as observed in the yellow tiles in the confusion matrices. This is an ideal result since our goal is to discriminate kinematical effects from dynamic pole structures. The confusion matrices show that our DNN models are capable of correctly recognizing the dynamic pole structures and will not mislabel them as triangle enhancements. To summarize, the confusion matrices reflected the ambiguity concerning the line shapes produced by the triangle mechanisms. Despite this, the analysis showed that the trained DNN models have no problem recognizing and classifying pole-based enhancements. Considering this and the fact that all trained DNN models obtained high accuracies during the training, testing, and validation stages, we move forward in applying them to perform an interpretation of the $P_\psi^N(4312)^+$ experimental data.

\subsection{Interpreting $P_\psi^N(4312)^+$}

\begin{table*}[ht!]
	\centering
	\renewcommand{\arraystretch}{1.5}{
		\caption{The final inference result using the LHCb data. The mean count out of 3000 bootstrapped inference amplitudes and their corresponding standard deviations obtained by the snapshot ensemble approach.}
		\label{tab:inference_sd}
		\begin{tabular*}{\textwidth}{@{\extracolsep{\fill}}>{\bfseries}lcccc}
			\hline\hline
			&\textbf{Triangle} & \textbf{1 pole in $[bt]$} & \textbf{1 pole in $[tb]$} & \textbf{1 pole each in $[bt]$ and $[bb]$} \\
			\hline\hline
			DNN 1 
			&$261\pm 39.4$ &$1930\pm166$ 
			&$0.248\pm0.454$ &$804\pm169$ \\
			DNN 2
			&$18.4\pm4.34$ & $1998\pm97.5$ 
			&$969\pm93.6$ &$14.5\pm7.23$ \\
			DNN 3
			&$0.653\pm1.29$ &$2960\pm22.2$ 
			&$0\pm0$ &$36.0\pm21.7$ \\
			\hline\hline
			DNN 4
			&$0\pm0$ &$2990\pm79.4$ 
			&$0\pm0$ &$14.2\pm79.4$ \\
			DNN 5
			&$0\pm0$ &$2660\pm274$ 
			&$0\pm 0$ &$337\pm274$ \\
			DNN 6
			&$0.436\pm3.77$ &$2999\pm9.73$ 
			&$0\pm0$ &$0.891\pm8.91$ \\
			\hline\hline
			DNN 7
			&$1\pm0$ &$2590\pm16.3$    
			&$0\pm0$ &$414\pm16.3$ \\
			DNN 8
			&$0\pm0$ &$2999\pm0.394$ 
			&$0\pm0$ &$1.06\pm0.394$ \\
			DNN 9
			&$0\pm0$ &$2870\pm14.3$ 
			&$0\pm0$ &$127\pm14.3$ \\
			\hline\hline
	\end{tabular*}}
\end{table*}

Recall that the design of our DNN takes in random energy points and the corresponding intensity values, in order to take advantage of the energy resolution and statistical uncertainties from the experimental data. It is expected that, due to the presence of error bars, the interpretation of the experimental data can also be spread out. This possible variation in the interpretation should be reflected in the final inference stage. We performed bootstrapping technique and generated 3,000 line shapes from the experimental data by randomly selecting a representative point from each energy bin, using a uniform distribution. For the intensity axis, we exploit the error bars present in the experimental data and similarly, randomly select a representative point for each intensity corresponding to our selected energy points.
We used snapshots of each DNN model to interpret the generated line shapes from the experimental data and calculated the mean and standard deviations of the predictions for each model. The standard deviation represents the confidence of the models in the interpretation as represented by the mean. The inference result is shown in Table \ref{tab:inference_sd}.

Notice that the triangle singularity interpretation for the $P_\psi^N(4312)^+$ is consistently ruled out by all of the trained DNN models. Although in the case of the DNN 1 inference, there is a small count of predictions for the triangle singularity, it is only around 8\% of the total inputted experimental line shapes and has a significant deviation from pole-based interpretation. The remaining trained DNN models interprets the same but with negligible number of predictions corresponding to triangle singularity. Recall that in the initial analysis done by LHCb \cite{LHCb:2019kea}, the triangle interpretation was ruled out due to some physical constraints, as discussed in section \ref{sec:trianglesingularity}. Here, in our result, the triangle interpretation is ruled out by using only pure line-shape analysis with the help of trained DNN. This conclusion becomes even more significant when considering that we deliberately incorporated hypothetical triangle mechanisms in order to achieve line shapes with prominent peaks that fit the experimental data. Despite this, the DNN still does not favor the triangle interpretation, allowing it to be ruled out as a viable explanation for the origin of $P_\psi^N(4312)^+$.

Furthermore, it is interesting to point out that the trained DNN models favor the pole-based interpretation, with a large percentage of predictions falling under the configuration of one pole in the [$bt$]-sheet. That is, the peak found below the $\Sigma_c^+\bar{D}^0$ threshold is most likely produced by an isolated pole in the second Riemann sheet. This result conforms with the expected line shape where the peak appears below the threshold when the pole is in the $[bt]$-sheet. Moreover, recall our observation in section~\ref{sec:uniformization} that pole-based enhancements generally have a narrower and asymmetric width compared to the triangle. This is one of the possible reasons why it is preferred by the DNN. Thus, naively speaking, a hadronic molecule interpretation corresponding to the label (1) classification may be inferred from our result as being the most favorable, which is indeed the case being argued by multiple studies \cite{Chen:2019bip,Cheng:2019obk,Du:2019pij}. However, it has already been shown in \cite{Santos:2023gfh} that such line shape is prone to ambiguous pole structure since it can also be produced by three poles distributed in each unphysical Riemann sheet. Addressing this ambiguity requires careful treatment where the contribution from the off-diagonal elements of the $S$-matrix must be included. A more rigorous probing of the pole structure of $P_\psi^N(4312)^+$ was recently addressed in \cite{Santos:2024bqr}.

Lastly, we also note that different probability distributions were employed in the bootstrapping process to generate the experimental line shapes used for inference. Remarkably, the results remained consistent across all cases. This demonstrates that the inference drawn from the trained DNN is robust and does not depend on how the inference dataset is generated from the experimental data. Therefore, to reiterate, our deep learning framework supports conclusions from previous analyses, indicating that the triangle singularity can indeed be ruled out as an explanation for the $P_\psi^N(4312)^+$ state.

\section{Conclusion and Outlook}\label{sec:conclusion}
In this study, we have demonstrated for the first time how a triangle singularity can be distinguished from pole-based enhancements through a deep learning framework. Using the formalism of triangle loops and uniformized poles, we produced training datasets that teach the DNN model to recognize line shapes arising from both kinematic and dynamic processes of a reaction. The trained DNN models were shown to provide reasonably high accuracy of discrimination between the two seemingly identical enhancements and also robust and consistent inference on the experimental data, despite the presence of experimental uncertainties. Using the trained DNN to analyze the line shape of the $J/\psi p$ invariant mass distribution near the $\Sigma_c^+\bar{D}^0$ threshold, the triangle singularity is consistently ruled out as the origin of the $P_\psi^N(4312)^+$ state.

The result of our analysis provides an independent validation to the analysis of LHCb that the triangle mechanism is an unlikely explanation of the enhancement just below the $\Sigma_c^+\bar{D}^0$ threshold \cite{LHCb:2019kea}. It must be pointed out that our present analysis does not claim that the pole-based enhancement for $P_\psi^N(4312)^+$ is the ultimate explanation for the observed signal. Other mechanisms should be included in the training dataset to make a more definitive description of the signal. Otherwise, with limited representation of possible mechanisms, the output of the DNN during the inference stage might be misleading. For example, our present analysis does not rule out the possibility that $P_\psi^N(4312)^+$ is due to a double triangle mechanism as discussed in \cite{Nakamura:2021qvy}. Inclusion of other mechanisms will be considered in our future investigations.

Moving forward, we aim to perform similar studies that cover a wider energy range if not the whole invariant mass distribution of the $J/\psi p$ and apply our framework on the $P_c(4440)^+$ and $P_c(4457)^+$ states, which we have already partly accomplished in \cite{Co:2024qnp}. This analysis would require working beyond two-channel treatments and investigating the possible coupling of other channels and their corresponding mechanisms \cite{Yan:2021nio, Yang:2022ezl,Yan:2022wuz}. Mixing of different processes that can enhance possible suppressed triangle mechanisms should also be considered so that the parameters used in the training can still be constrained physically. Similar studies may also be done to evaluate the triangle singularity interpretation for various other potential candidates \cite{Guo:2019twa}. Moreover, we recommend that further studies be done to establish a more definitive interpretation for the $P_\psi^N(4312)^+$ state such as accommodating more sophisticated pole configurations, probing even distant Riemann sheets, and settling the ambiguity problem revealed in \cite{Santos:2023gfh, Santos:2024bqr}.

In closing, we would like to remark that our study is the first-ever known case of the development of a model-selection framework in hadron physics. The concept of model selection is mostly used in the fields of statistics and machine learning, wherein the most appropriate model is often selected among a set of candidates following some measure or criteria. This is done to obtain the best model that accurately explains experimental data while providing reliable predictions \cite{Zucchini200041}. In our case, we have developed a framework that can select the most favorable mechanism to interpret a measured enhancement, while eliminating those misleading ones. For instance, the pole-based interpretation was selected as the most likely origin of the $P_\psi^N(4312)^+$ signal by the trained DNN. Consequently, the triangle singularity interpretation was then ruled out for this state. It must be noted that model selection is only suggestive in nature and never completely confirmatory \cite{Ding8498082}. Hence, it still demands supporting studies that can provide in-depth analyses and reasonable conclusions. Although our model-selection framework is currently limited to the chosen mechanisms used to train the DNN, its effective demonstration proves that it can serve as a valuable tool in hadron physics for classifying exotic hadron candidates. The novel framework offers an alternative approach to studying line shapes and supplements the standard fitting methods. It can be extended to include more theoretical models and potentials used in hadron spectroscopy. Exotic hadrons may then be analyzed using this framework in order to gain stronger intuition on the origin of their signals.

\begin{acknowledgments}
DAOC acknowledges the support of the DOST-SEI Undergraduate Merit Scholarship Program. VAAC acknowledges the support of DOST-ASTHRDP. The authors acknowledge the helpful suggestions of Leonarc Michelle Santos and Mao-Jun Yan. DLBS acknowledges the insightful discussions with Prof. Tetsuo Hyodo and Prof. Atsushi Hosaka during the long-term workshop on HHIQCD2024 at the Yukawa Institute for Theoretical Physics (YITP-T-24-02).
\end{acknowledgments}

\bibliography{mybib}

\onecolumngrid
\newpage
\appendix
\section{Sample Datasets}
\label{sec:appendix}
\begin{figure*}[h!]
	\centering
	\includegraphics[width=0.65\linewidth]{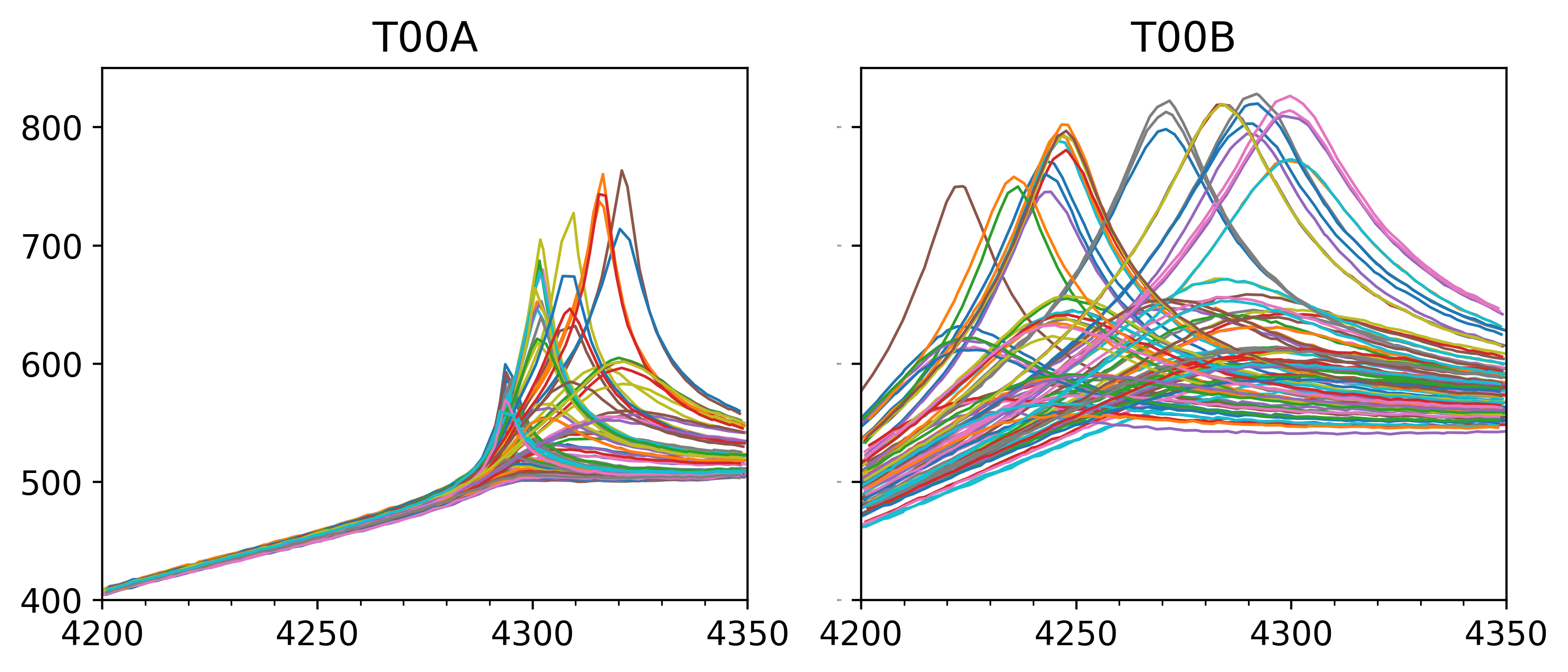}
	\caption[]{Sample dataset of constructed triangle singularity line shapes.}\label{fig:Sample_TS}
\end{figure*}

The crucial step in the line shape analysis via machine learning is the generation of the training dataset. As discussed in the main text, two triangle mechanisms are used in this study. These triangle line shapes are shown in Fig.~\ref{fig:Sample_TS}. The T00A represents the triangle diagram shown in Fig.~\ref{fig:TSloopA} while T00B corresponds to Fig.~\ref{fig:TSloopB}. The appearance and location of peaks for the triangle line shapes can already be inferred based on the masses involved in the internal triangle diagram (see section \ref{sec:trianglesingularity}). The variation in the prominence of the peaks is attributed to the different decay width and form factor used. Some of these line shapes do not necessarily resemble the actual $P_{\psi}^{N}(4312)^+$ signal but they shall be considered to help the DNN learn the general features of the triangle line shapes.

\begin{figure*}[h!]
	\centering
	\includegraphics[width=0.9\linewidth]{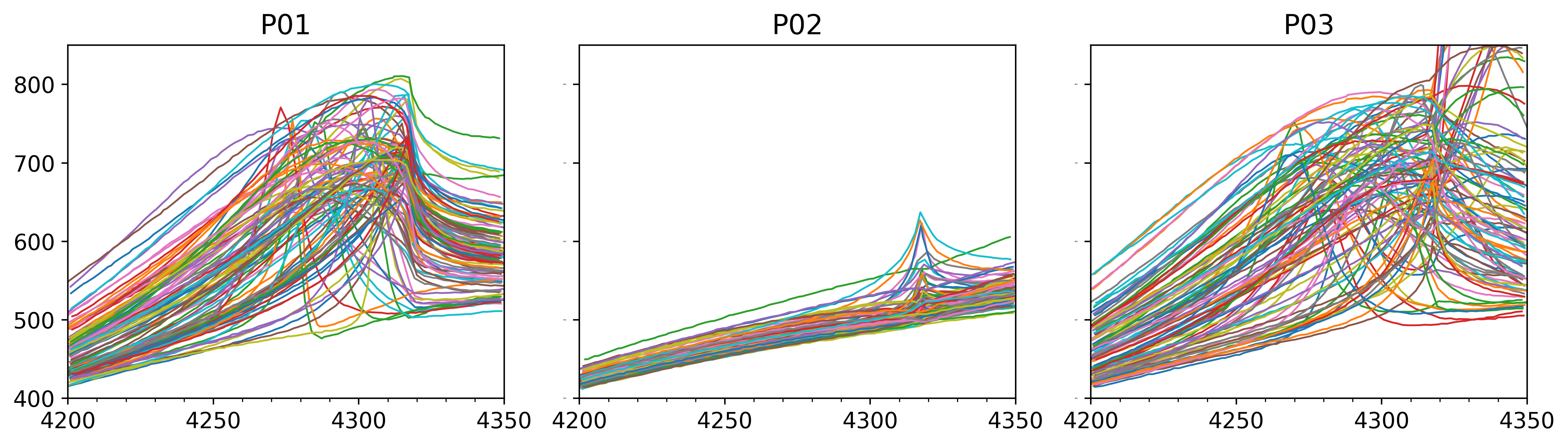}
	\caption[]{Sample dataset of constructed pole-based line shapes.}\label{fig:Sample_Poles}
\end{figure*}

On the other hand, Fig.~\ref{fig:Sample_Poles} shows the three pole configurations considered in the study. The P01 and P02 training sets are the label used for line shapes with an isolated pole in the $[bt]$-sheet and $[tb]$-sheet, respectively. The general features of these line shapes, specifically the appearance of peak relative to the second threshold ($\Sigma_c^+\bar{D}^0$), is shared by the datasets. Notice that the peak is relatively weak in P02 compared to the other classifications, which is also demonstrated in Fig.~\ref{fig:btbbtbpeaks}. The suppressed strength of an isolated pole in the $[tb]$-sheet is due to the fact that it is located in the farthest Riemann sheet among the unphysical sheets considered. Furthermore, the peak is also restricted by the unitarity of the $S$-matrix between the two thresholds, suppressing the signal even further.
Lastly, the P03 dataset has the most non-trivial line shapes among the different interpretations considered in this work. Recall that two poles are used in P03, one in $[bt]$ and the other one in $[bb]$. In some cases, these poles can produce two peaks in the line shapes and is easily distinguishable from the other classifications. However, with this configuration, there are cases when the $[bt]$ and the $[bb]$ poles are both below the second threshold. In this situation, only one peak is observed associated with the $[bt]$ pole while the $[bb]$ pole serves to lessen the cusp around the second threshold. In fact, when these poles have exactly the same position, the peak will have a symmetric shape despite the presence of the second threshold. This last pole-based classification is expected to produce line shapes that is quite similar with the triangle shapes. This observation makes the line shape classification a non-trivial task without the use of an automated pattern recognition method such as machine learning.

\section{Uncertainty Estimation}
\label{sec:appendixx}

\begin{figure*}[h!]
	\centering
	\includegraphics[width=\linewidth]{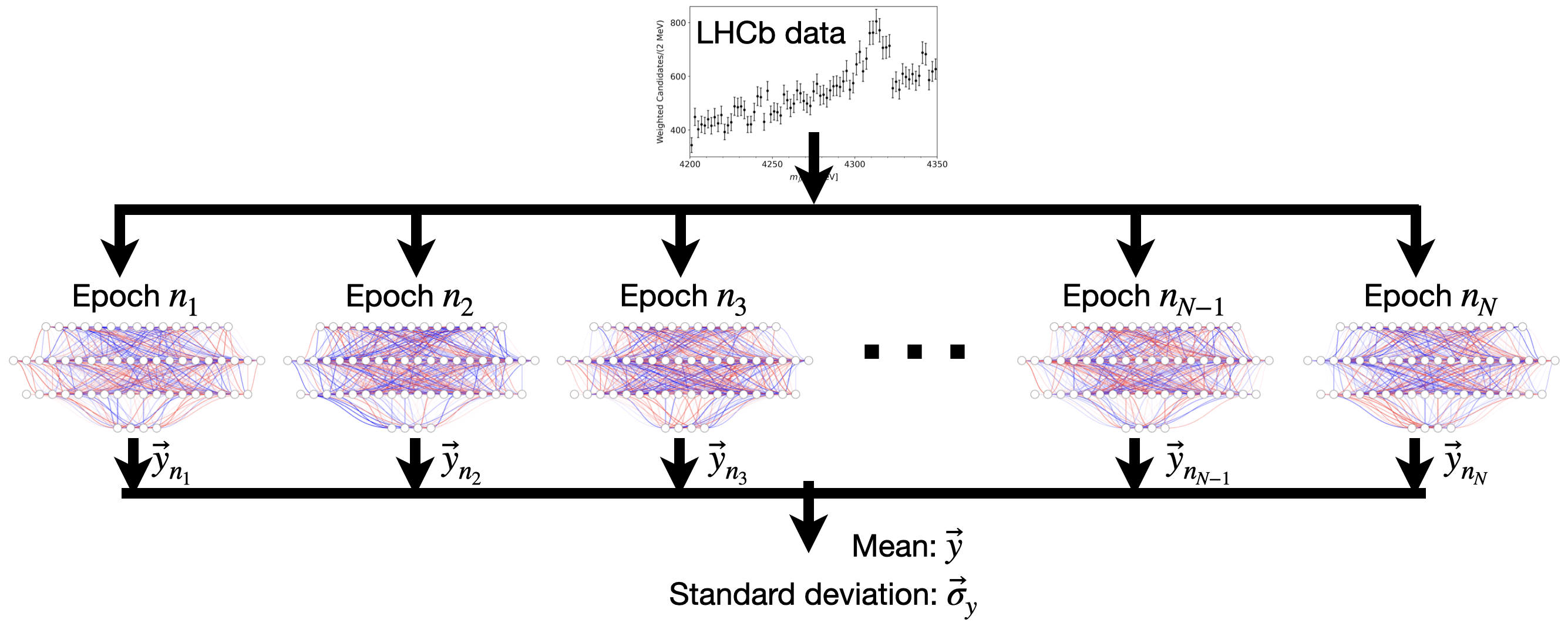}
	\caption[]{Schematic diagram of the snapshot ensemble method to estimate the uncertainty of the DNN's output.}\label{fig:snapshot}
\end{figure*}

In this appendix, we briefly describe how the snapshot ensemble is used to estimate the uncertainty of our trained DNN. During training, we saved the state of our DNN models from epoch 950 to epoch 1050, creating a snapshot ensemble of 100 different DNN instances. This ensemble is utilized to estimate the uncertainty of the inference results given the experimental data. Our approach is schematically outlined in Fig.~\ref{fig:snapshot}. We first generated 3000 inference samples by bootstrapping the given experimental data. These 3000 samples were obtained using a uniform distribution, with extreme values marked by the error bars and energy uncertainty present in the experimental data. These samples were then fed directly into the 100 snapshot states of the DNN, and the number of classification outputs was counted. From the distribution of the DNN's outputs, we can estimate the mean and standard deviations for each classification. This method provides a comprehensive measure of uncertainty, which is crucial for understanding the reliability of the model's predictions and for making informed decisions based on the inference results.

\end{document}